\documentclass[preprint,12pt]{elsarticle}

\usepackage{amssymb}
\usepackage{amsmath}
\usepackage{soul}
\usepackage{url} 
\usepackage{hyperref}
\usepackage{graphicx}

\usepackage{ulem} % put a line crossing a letter

\usepackage{lineno}
\usepackage{multirow}
\usepackage{caption}
\usepackage{subcaption}  
\usepackage{textcomp}
\usepackage[T1]{fontenc}
\usepackage[utf8]{inputenc}

\journal{NIM-A}

\begin{document}

\begin{frontmatter}

%% Title, authors and addresses

%% use the tnoteref command within \title for footnotes;
%% use the tnotetext command for theassociated footnote;
%% use the fnref command within \author or \affiliation for footnotes;
%% use the fntext command for theassociated footnote;
%% use the corref command within \author for corresponding author footnotes;
%% use the cortext command for theassociated footnote;
%% use the ead command for the email address,
%% and the form \ead[url] for the home page:
%% \title{Title\tnoteref{label1}}
%% \tnotetext[label1]{}
%% \author{Name\corref{cor1}\fnref{label2}}
%% \ead{email address}
%% \ead[url]{home page}
%% \fntext[label2]{}
%% \cortext[cor1]{}
%% \affiliation{organization={},
%%            addressline={}, 
%%            city={},
%%            postcode={}, 
%%            state={},
%%            country={}}
%% \fntext[label3]{}

\title{Characterization of the optical model of the T2K 3D segmented plastic scintillator detector} %% Article title

%% use optional labels to link authors explicitly to addresses:
%% \author[label1,label2]{}
%% \affiliation[label1]{organization={},
%%             addressline={},
%%             city={},
%%             postcode={},
%%             state={},
%%             country={}}
%%
%% \affiliation[label2]{organization={},
%%             addressline={},
%%             city={},
%%             postcode={},
%%             state={},
%%             country={}}

\author[20]{S.~Abe}
\author[1, 10]{I.~Alekseev}
\author[12]{T.~Arai}
\author[2]{T.~Arihara}
\author[19]{S.~Arimoto}

\author[15]{N.~Babu}
\author[3]{V.~Baranov}
\author[4]{L.~Bartoszek}
\author[21]{L.~Berns}
\author[15]{S.~Bhattacharjee}
% ...
% The "\note" macro will give a warning: "Ignoring empty anchor..."
% you can safely ignore it.

%%\affiliation{organization={},
%%            addressline={}, 
%%            city={},
%%            postcode={}, 
%%            state={},
%%            country={}}
\affiliation[1]{organization={Lebedev Physical Institute of the Russian Academy of Sciences}, 
                addressline={53 Leninskiy Prospekt}, 
                city={Moscow}, 
                postcode={119991}, 
                country={Russia}}
\affiliation[2]{organization={Tokyo Metropolitan University, Department of Physics}, 
                city={Tokyo}, 
                country={Japan}}
\affiliation[3]{Joint Institute for Nuclear Research, Dubna, Moscow Region, Russia}
\affiliation[4]{University of Colorado, Boulder, Colorado 80309 USA}
% ...
\affiliation[10]{Institute for Nuclear Research of RAS, Moscow, Russia}
% ...
\affiliation[12]{University of Tokyo, Department of Physics, Tokyo, Japan}
% ...
\affiliation[15]{Louisiana State University, Department of Physics and Astronomy, Baton Rouge, Louisiana, USA}
% ...
\affiliation[19]{Kyoto University, Department of Physics, Kyoto, Japan}
\affiliation[20]{University of Tokyo, Institute for Cosmic Ray Research, Kamioka Observatory, Kamioka, Japan}
\affiliation[21]{Tohoku University, Faculty of Science, Department of Physics, Miyagi, Japan}
% ...
\cortext[cor1]{Corresponding author}

%% Abstract
\begin{abstract}
The magnetised near detector (ND280) of the T2K long-baseline neutrino oscillation experiment has been recently upgraded aiming to satisfy the requirement of reducing the systematic uncertainty from measuring the neutrino-nucleus interaction cross section, which is the largest systematic uncertainty in the search for leptonic charge-parity symmetry violation.
A key component of the upgrade is SuperFGD, a 3D segmented plastic scintillator detector made of approximately 2,000,000 optically-isolated $1~\text{cm}^{3}$ cubes. It will provide a 3D image of GeV neutrino interactions by combining tracking and stopping power measurements of final state particles with sub-nanosecond time resolution. 
The performance of SuperFGD is characterized by the precision of its response to charged particles as well as the systematic effects that might affect the physics measurements.

Hence, a detailed Geant4 based optical simulation of the SuperFGD building block, i.e. a plastic scintillating cube read out by three wavelength shifting fibers, has been developed and validated with the different datasets collected in various beam tests. In this manuscript the description of the optical model as well as the comparison with data are reported.  
\end{abstract}

%% Keywords
\begin{keyword}
%% keywords here, in the form: keyword \sep keyword
Particle detector, plastic scintillator, optical simulation, neutrino oscillations, T2K
%% PACS codes here, in the form: \PACS code \sep code

%% MSC codes here, in the form: \MSC code \sep code
%% or \MSC[2008] code \sep code (2000 is the default)

\end{keyword}

\end{frontmatter}

\flushbottom
%%\linenumbers

\section{Introduction }

\label{sec:intro}

The T2K long-baseline neutrino oscillation experiment \cite{T2KExperiment} searches for leptonic CP violation by looking for an asymmetry in the oscillation probability between neutrinos and antineutrinos.
Protons are accelerated up to about 30~GeV and impinging on a graphite target, producing mainly pions and kaons. Then, these particles decay and produce a (anti)neutrino beam with a characteristic energy spectrum peaked around 600 MeV.
Neutrinos are detected by two different detector complexes:
a near detector (ND280), positioned at about 280~m from the production target, and a far detector (Super-Kamiokande), at a distance of 295~km where the probability that neutrinos have undergone flavour oscillations is maximum.
While the water-Cherenkov far detector is gigantic to collect as much data as possible, ND280 has to precisely detect neutrino interactions to constrain systematic uncertainties such as those related to the neutrino flux and the neutrino-nucleus interaction cross section, that made the largest contribution.  

ND280 consists of a neutrino beam monitor and a magnetised detector complex \cite{T2K:2019tech} that recently underwent a major upgrade to allow for a more precise characterisation of neutrino-nucleus interactions. 
In order to augment the performance of the already existing plastic scintillator fine-grained detectors (FGD1 and FGD2), 
made of 2 m long scintillating bars as neutrino active target, whose horizontal and vertical deployment is alternated along the neutrino beam direction,
a novel detector, named SuperFGD, has beed constructed and installed in 2023 \cite{nd280upgrade-press-release}.    
Its concept is described in Ref.~\cite{Sgalaberna:2017khy}.
It consists of 1956864 1 $\text{cm}^3$ plastic scintillator cubes with 55888 readout channels. The scintillator composition is polystyrene doped with 1.5\% paraterphenyl (PTP) and 0.01\% POPOP. Each cube is optically isolated and is read out by three orthogonal wavelength shifting fibers (Kuraray Y11 double cladding \cite{kuraray-catalogue-wls}) that guide the scintillation light towards Hamamatsu S13360-1325PE multi-pixel photocounters (MPPC) \cite{hamamatsu:mppc}. 
3D images of neutrino interaction final states can be obtained by combining the signatures in the three 2D readout views.
Thanks to its highly-segmented geometry, SuperFGD is able to reconstruct charged particles produced isotropically and protons down to about 300-350 MeV/c. This is a major improvement with respect to FGD1 and FGD2. Moreover, with its higher scintillation light yield, SuperFGD will provide simultaneously an improved particle identification. It is expected to largely reduce the contamination from $\gamma \rightarrow e^+e^-$ in the sample of selected $\nu_e$ by precisely measuring the stopping power (dE/dx), and to detect neutrons 
and reconstruct their time of flight thanks to the sub-nanosecond single-cube time resolution.

It is crucial to estimate properly the performance of SuperFGD, which depends on the understanding and the characterisation of its response to charged particles as well as to the identification of possible systematic effects.
In the past few years, multiple SuperFGD prototypes have been manufactured and exposed to particle test beams. A first small prototype consisting of 5 $\times$ 5 $\times$ 5 cubes was put in test beam at CERN in 2018 \cite{sfgd-2018}, providing promising measurement of the scintillation light yield and the crosstalk, i.e., the light leakage from a scintillator cube to an adjacent one. A larger prototype consisting of 24 $\times$ 8 $\times$ 48 cubes was later manufactured, assembled and tested again in a test beam at CERN in 2019, which was reported in \cite{Blondel:2020sfgd}. A more accurate study of the light crosstalk was carried out by looking at highly-ionizing protons stopping inside the detector. A sub-nanosecond time resolution of the detector was estimated from the time information simultaneously collected in multiple cubes and, later, confirmed in Ref.~\cite{TR_Alekseev:2022jki} with a controlled laser source. 
Another small prototype has been tested in Tohoku University in Japan. 
The light yield non-uniformity as a function of the charged particle position within a single plastic scintillator cube was measured. 
Furthermore, a stand-alone measurement of the attenuation property of the selected WLS fibers has been done. The later two experiments were carried on recently and their description and main results are reported in this article.

Based on the measurements of the SuperFGD prototypes, a Geant4-based optical simulation has been developed and fine-tuned by this work. The optical simulation will become an important toolbox for gaining a thorough understanding on the optical processes happened in a single SuperFGD detection unit, and as a flexible simulation structure for studying further optimization of the detector configuration. The simulation was fine-tuned based on the datasets collected with the different prototypes. The detailed description of the optical simulation and the comparison with the available data is reported.

\section{Datasets}
\label{sec:dataset}

\subsection{2018 protoype beam test}
\label{sec:2018_data}
A small SuperFGD prototype made of 5 $\times$ 5 $\times$ 5 cubes has been tested with a charged particle beam in 2018 at the T10 area of the CERN Proto-Synchrotron (PS)~\cite{sfgd-2018}. Each cube was read out by three orthogonal 1.3 m long  Kuraray Y11 double cladding wavelength shifting (WLS) fibers of 1 mm diameter~\cite{kuraray_y11}. One end of the fiber was coupled to a Hamamatsu S12571-025C multi-pixel photon counter (MPPC)~\cite{hamamatsu_s13360-1325cs}, with a nominal photodetection efficiency (PDE) of 35\%, by a 3D-printed black connector, where the coupling was ensured by enhancing the contact between the MPPC and the WLS fiber end with a soft elastic foam placed inside the connector at the base of the ceramic-type MPPC. The other end of the WLS fiber, about 7~cm far from the closest scintillator cube, was covered with aluminized reflective paint to increase the light yield. Two trigger scintillator counters were put before and after the prototype to select minimum ionizing particles (MIPs), ensuring consistent energy deposition of the charged particles within the cube volume. 

On average, a light yield of 83.1 p.e./MIP/2 channels (41.6 p.e./MIP/channel) was measured. The light yield data of this prototype beam test were taken as the reference in this work. Without loss of generality, we focus on the light yield of a single fiber channel in the following context.

\subsection{2019 prototype beam test}
\label{sec:2020_data}
In 2019 a larger prototype made of 24 $\times$ 8 $\times$ 48 cubes was assembled and tested with a charged particle beam at the CERN PS facility~\cite{Blondel:2020sfgd}. In total 1728 WLS fiber channles, of either 8 cm or 24 cm in length depending on the read out view, were used. One end of the fiber was coupled to MPPCs by a 3D-printed connenctor in the same way as in the prototype in Sec.~\ref{sec:2018_data}. The other end of the fibers were left unpolished and open without any coating. Three types of Hamamatsu MPPCs, with nominal PDE ranging from 25\% to 35\%, were used to equip the prototype.
Thanks to the particle identification obtained by the beamline instrumentation, the light crosstalk between adjacent cubes could be precisely measured from a sample of higher light yield events such as highly-ionizing protons stopping in the prototype, while a more accurate time resolution was derived simultaneously with this prototype thanks to its larger active volume and the higher statistics.
Thus, the simulated light crosstalk and time resolution of this work will be compared to the data measured with this prototype.  

\subsection{Tohoku beam test}
\label{sec:tohoku-beam-test}
\begin{figure}[hbtp] 
	\begin{center}
		\includegraphics[width=0.65\textwidth]{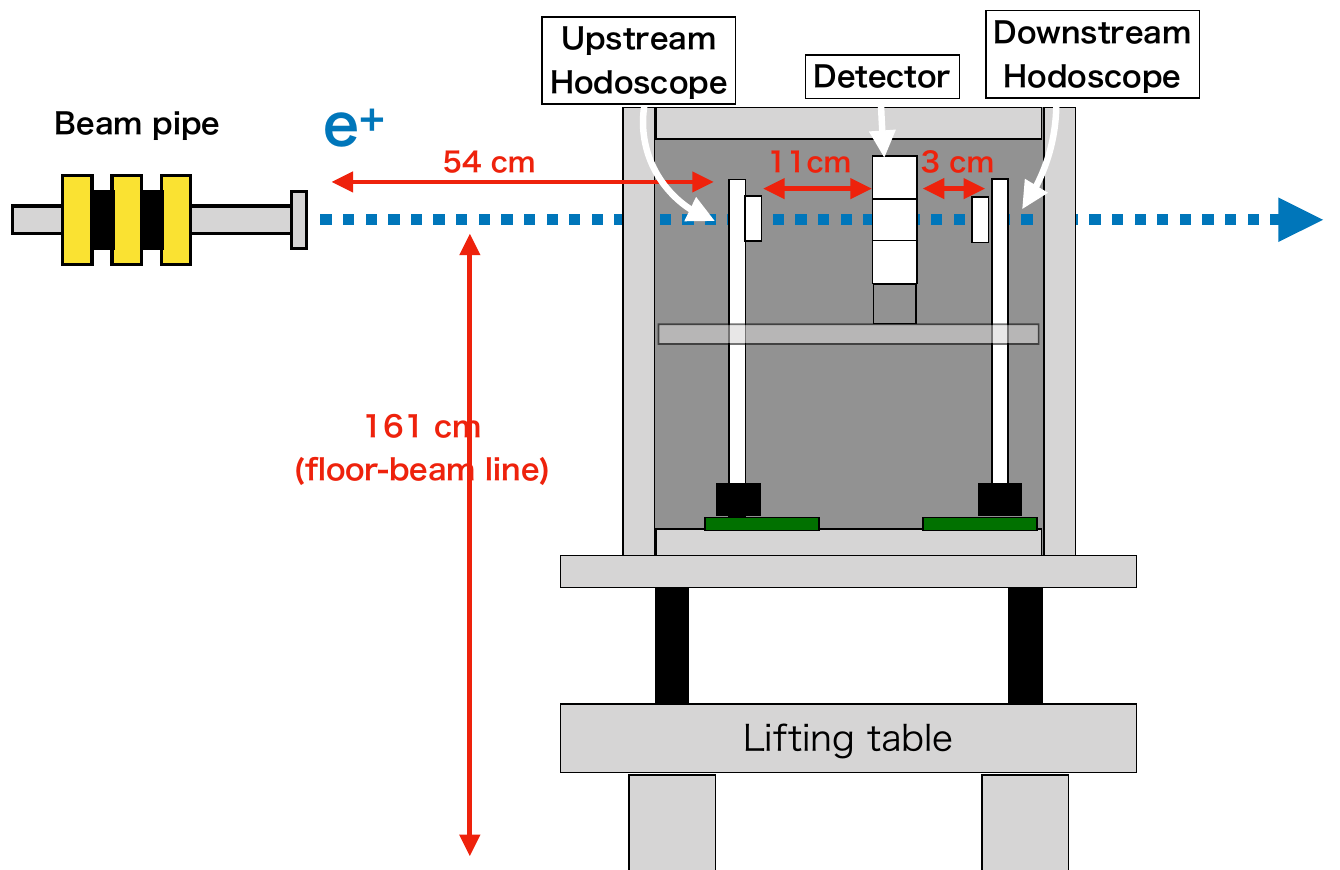}
		\caption{Setup of the Tohoku beam test.}
		\label{fig:setup}
	\end{center}
\end{figure}

Another beam test was performed at the GeV-$\gamma$ experimental hall in the research center for ELectron PHoton science (ELPH), Tohoku University\footnote{Currently, Research Center for Accelerator and Radioisotope Science (RARiS), Tohoku University}, to characterize the non-uniformity of a single SuperFGD cube as a function of the interacting point of a traversing charged particle. 
The schematic drawing of the beam test setup is shown in Fig.~\ref{fig:setup}.
A 500 MeV/c positron beam is produced by accelerators in ELPH.
The scintillator prototype is put on the beam axis. We utilizes four hodoscopes located at the upstream and downstream of the scintillator cube sample to precisely measure the beam position.

\subsubsection{Beamline}
First, an electron beam is accelerated up to 200~MeV at a maximum by a linear accelerator.
Then, the electron beam is injected into a synchrotron called Booster-STorage (BST), and accelerated up to 1.3~GeV at a maximum.
By a carbon wire in orbit of the accelerated electrons, high energy $\gamma$-rays are generated via bremsstrahlung.
Electron-positron pairs are produced from the $\gamma$-rays at a tungsten target.
A bending magnet analyzes momentum of the electron or the positron.
We used the 500 MeV/c positron beam.
The width of the beam is about 7 mm in both x and y axes and monitored thought the test. The beam rate is about 3~kHz with a duty cycle of 0.63 at the maximum.

\subsubsection{Scintillator prototype}
We evaluated the light yield and optical crosstalk. The first setup, consisting of a single cube, focuses the characterisation of the observed scintillation light yield as a function of the position of the beam particle. 
The second setup, consisting of a single layer of 3 $\times$ 3 cubes is used to evaluate the optical crosstalk probability to the adjacent channels. About 30\,cm long WLS fibers (Kuraray Y-11 (200)M) were used. The WLS fibers are inserted in each cube along the three orthogonal directions and coupled with MPPCs, Hamamatsu S13360-1325PE, which are the same type as those of the SuperFGD detector. The signal from the MPPCs is digitized by a front-end readout module that uses the EASIROC ASIC \cite{CALLIER20121569}. In each measurement the scintillator cube is placed in the center of the beamline.

\subsubsection{Hodoscopes}
Each hodoscope consists of arrayed 16 scintillating fibers, Kuraray SCSF-78SJ.
The fiber has square shape with the cross section of 1.5 $\times$ 1.5~mm$^2$, which is manually painted with TiO$_2$-based reflector of about 0.1~mm thickness. A pair of hodoscopes are perpendicularly stacked for the tracking. Scintillation light is detected by a 4 $\times$ 4~mm$^2$ MPPC array, S13361-3050AE-04 produced by Hamamatsu Photonics K.K, coupled with the fiber bundle.
The MPPCs are connected to the EASIROC readout module.

\subsubsection{Data acquisition}
The beam test at ELPH was performed in two periods, June 1-3 and November 13-15, 2018.
Since the statistics and quality of the data collected during the first period were not good enough for analysis due to issues with the data acquisition system, the analysis reported in this paper focuses on the second period data.
The downstream hodoscopes are used as trigger.
When there are hits in any of the 32 downstream hodoscope channels, data of all the hodoscopes and the scintillator cube sample are recorded.

\subsubsection{Event selection}
The hodoscopes are used to select events with a straight track perpendicular to the scintillator cube surface.
We defined channels greater than 2.5 photo-electron (p.e.) observed by the hodoscopes as hits. Then, events are selected if there are hits in the given upstream and corresponding downstream segments. We took 300~events of data at least, which provides sufficient statistics with less than 5\% uncertainty. Gain (ADC to p.e. conversion constant) calibration is performed by fitting the several p.e. peaks for the each single MPPC independently.

\subsubsection{Results}

\begin{figure}[hbtp] 
	\begin{center}

		\includegraphics[width=0.35\textwidth]{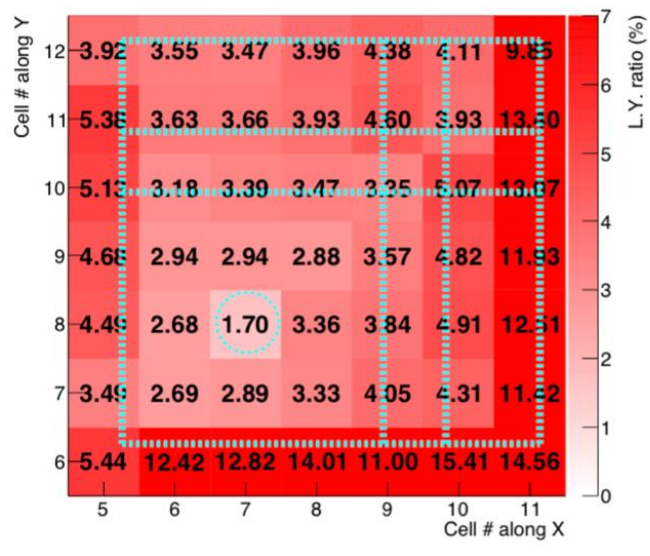}
		\caption{Result of optical crosstalk.}
		\label{fig:elph_crosstalk}
	\end{center}
\end{figure}

As expected, we found larger amount of light for the trajectory segments close to the fiber position. A detailed map was shown later in Sec.~\ref{sec:optsim;4} in Fig.~\ref{fig:nonuni_MC}. The overall light yield non-uniformity within a single cube, defined by the standard deviation, is respectively 6.7\%, 6.4\% in X and Y fiber channels (transverse to the beam direction) and 18.8\% in Z fiber channel (parallel to the beam direction). 
Figure~\ref{fig:elph_crosstalk} shows the result of the crosstalk probability observed for the trajectory segments, where the cross section of a single cube has been segmented into 7 $\times$ 7 cells showing.
The dashed line shows the surface of the cube covered by the hodoscopes, and the positions of the three holes inserted by the WLS fibers.
Higher probability is found if the trajectory is close to the reflective boundary with the neighbouring cube or close to the WLS fiber hole. On the other hand, low crosstalk probability between the diagonal cubes are found.

\subsection{WLS fiber attenuation measurements}
\label{fiber_att}
\begin{figure}[h!]
    \centering
            \includegraphics[width=0.6\linewidth]{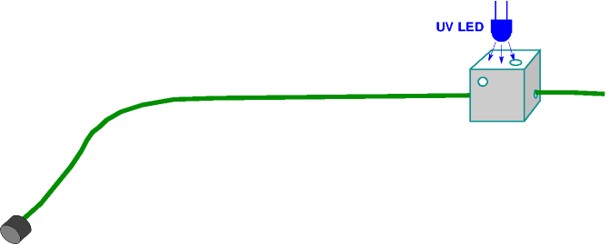}
            \caption{The set-up for measurement of the attenuation parameters of WLS fibers. }
        \label{fig:uv_led_test_setup}
\end{figure}

\begin{figure}[h!]
    \centering
            \includegraphics[width=0.6\linewidth]{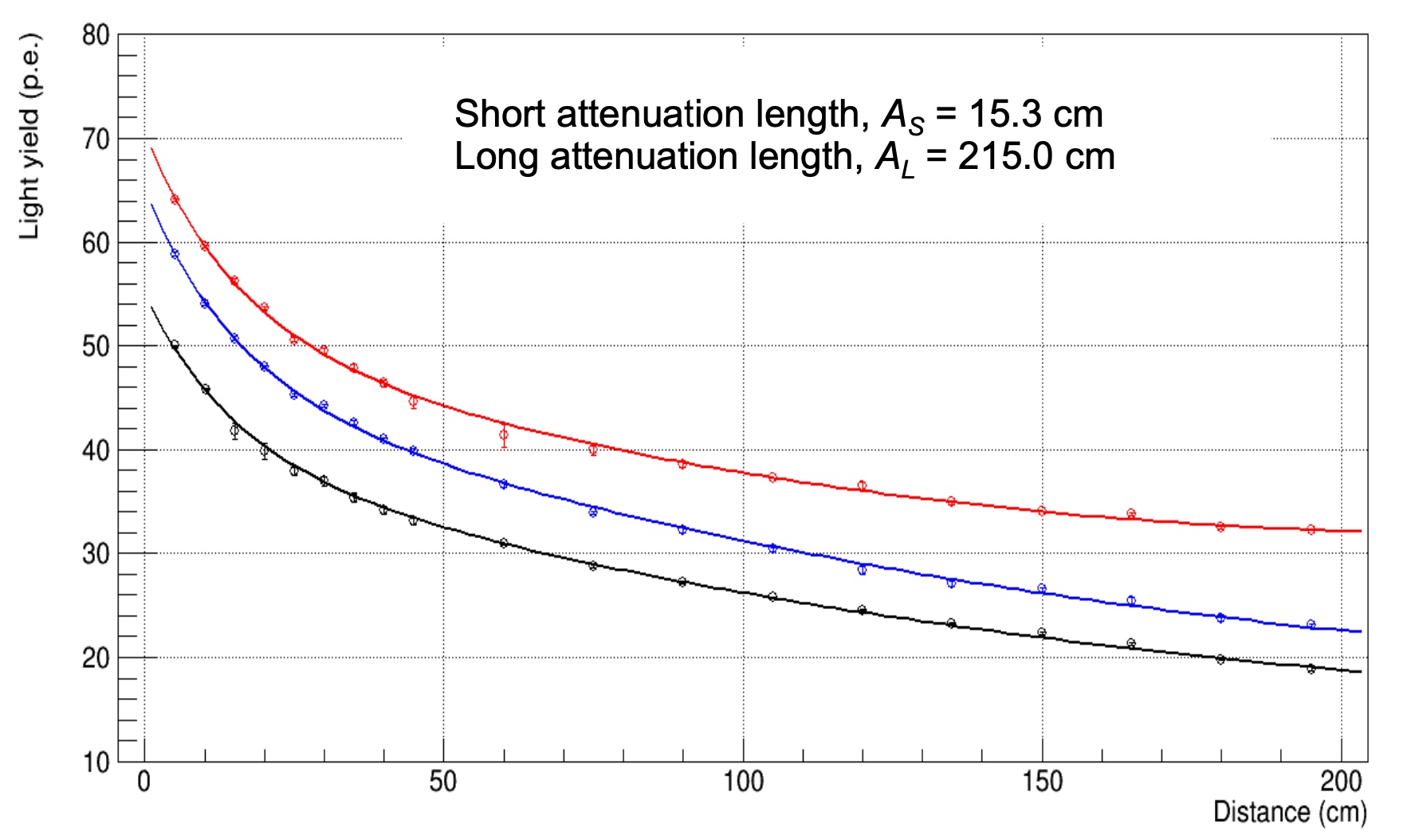}
            \caption{Light yield as a function of the distance from the illuminated cube to the MPPC. Red curve - the fiber end is polished and covered by a reflective paint; blue curve - fiber end is polished; black curve -  fiber end is cut under 45 degrees and painted by a black paint.  }
        \label{fig:fiber_attenuation}
\end{figure}

In this test, the light attenuation property of a WLS fiber was characterized.
Kuraray Y11 (200) S-type  double cladding WLS fibers of  1 mm diameter  and 1.3 m length   were used  in the tests. This same type of fiber was used in the other protypes as well as in the final SuperFGD detector.
The parameters of these fibers were measured using the setup configuration shown in Fig.~\ref{fig:uv_led_test_setup}. 
Pulsed collimated UV LED illuminated $1 \times 1 \times 1$ cm$^3$ scintillator cube through a side where the reflector  was removed. The LED flash intensity was adjusted to  approximately correspond to the light yield of a MIP.  The far end of a fiber inserted in  a hole was connected to an Hamamatsu MPPC S13360-1325CS. Fiber parameters were measured for three different cases: 1) polished fiber end covered by  a reflective paint; 1) polished fiber end without a reflector;  3) fiber end cut at approximately 45 degrees and painted with a black paint. 
Experimentally, the fiber response can be fit by the sum of exponential functions. Two approaches were used to fit the experimental data.\\
1. The light output of a fiber is parameterized by:
\begin{equation}
LY (x) = LY_{0S}\times \exp{(-x/A_S)} +  LY_{0L}\times \exp{(-x/A_L)}
 \label{eq:att_no_reflection}   
\end{equation}
where $A_{S}$ is the short attenuation length, $A_{L}$ is the long attenuation length,  and $LY (0) = LY_{0S} + LY_{0L}$. Using eq.~(\ref{eq:att_no_reflection}), the attenuation data are well fit with  $A_{S} = 24.82 \pm  0.19$ cm, $A_{L} = 619.4 \pm  8.7$ cm (case 1);  $A_{S} = 24.24 \pm  0.35$ cm, $A_{L} = 303.6 \pm  6.9$ cm (case 2); $A_{S} = 17.94 \pm  0.15$ cm, $A_{L} = 280.4 \pm  1.9$ cm (case 3). \\
2. An additional parameter, the reflection coefficient $R$, was added to the fitting function:

\begin{equation}
\begin{split}
LY (x) = &LY_{0S}\times \exp{(-\frac{x}{A_S}}) +  LY_{0L}\times \exp{(-\frac{x}{A_L})} \\
+& R \times [LY_{0S}\times \exp{(-\frac{2L - x}{A_S})} +  LY_{0L}\times \exp{(- \frac{2L - x}{A_L})}]
 \label{eq:att_with_reflection}
\end{split}
\end{equation}
where $L$ is the total length of the fiber. Results of  measurements and attenuation parameters are shown in Fig.~\ref{fig:fiber_attenuation}.
As seen from Fig.~\ref{fig:fiber_attenuation}, the light yield is higher when the fiber end is polished and covered by a reflector. Consistent light yield ratio ($LY_{0S}/LY_{0L}$) was observed as about 0.41, while the attenuation length  values are close to $A_S \simeq 15.13$ cm, $A_L \simeq 215.0 $ cm, with $R = 0.77,~ 0.26,~ 0.19$ for cases 1,2,3, respectively. 

\subsection{Measurement of time resolution}
\label{time_reso}

A time resolution study of the SuperFGD cubes has been reported in~\cite{TR_Alekseev:2022jki}. 
First, the beam test data described in Sec.~\ref{sec:2020_data} have been analysed and discussed, obtaining
a time resolution of 0.97 ns with the MIP events, including the systematic uncertainty of about 0.7 ns from the readout electronics. The time calibration and correction for time-walk effects was implemented. 
Furthermore, a 3 $\times$ 3 SuperFGD cube prototype was tested using a laser and read out by a wide-bandwidth oscilloscope, that the systematic error contributed by the electronics is negligible in this case. The wavelength of the laser was chosen to be the one that matches the absorption spectrum of the wavelength shifting doping of the plastic scintillator, to ensure a sufficiently high efficiency of the scintillating light production. Different light yield were generated by varying the intensity of the laser source. Then, a function between the light yield and time resolution has been obtained.
With negligible electronic systematic errors, a time resolution of 0.68 ns was measured at the beam test light yield. 

\section{Optical simulation}
\label{sec:optsim}
A detailed optical simulation of a single SuperFGD cube was developed using the Geant4 software \cite{AGOSTINELLI2003250}, in which all the optical processes were simulated in detail. Focusing on several important detector optical properties, such as the light yield, the cube-to-cube crosstalk, the time resolution and the light yield uniformity as a function of the interacting position of a traversing charged particle within a single cube, we tuned and compared the Monte Carlo (MC) results with multiple beam test datasets of SuperFGD prototypes, described in the previous sections, aiming to build a general simulation framework that can be used to quantify the detector systematic uncertainties of SuperFGD in detail. Furthermore, the framework has the potential to be used in the future for improving designs of analogous detectors.

In this section, we briefly introduce the software implementation including the geometry and the optical model of a single SuperFGD cube. Then, in order to fine tune the model parameters, the key optical observables mentioned above are compared to the data collected by exposing SuperFGD prototypes to test beams.   

\subsection{Geometry and optical model}
\label{sec:optsim;2}
\begin{figure}[h!]
    \centering
    \includegraphics[width=0.7\linewidth]{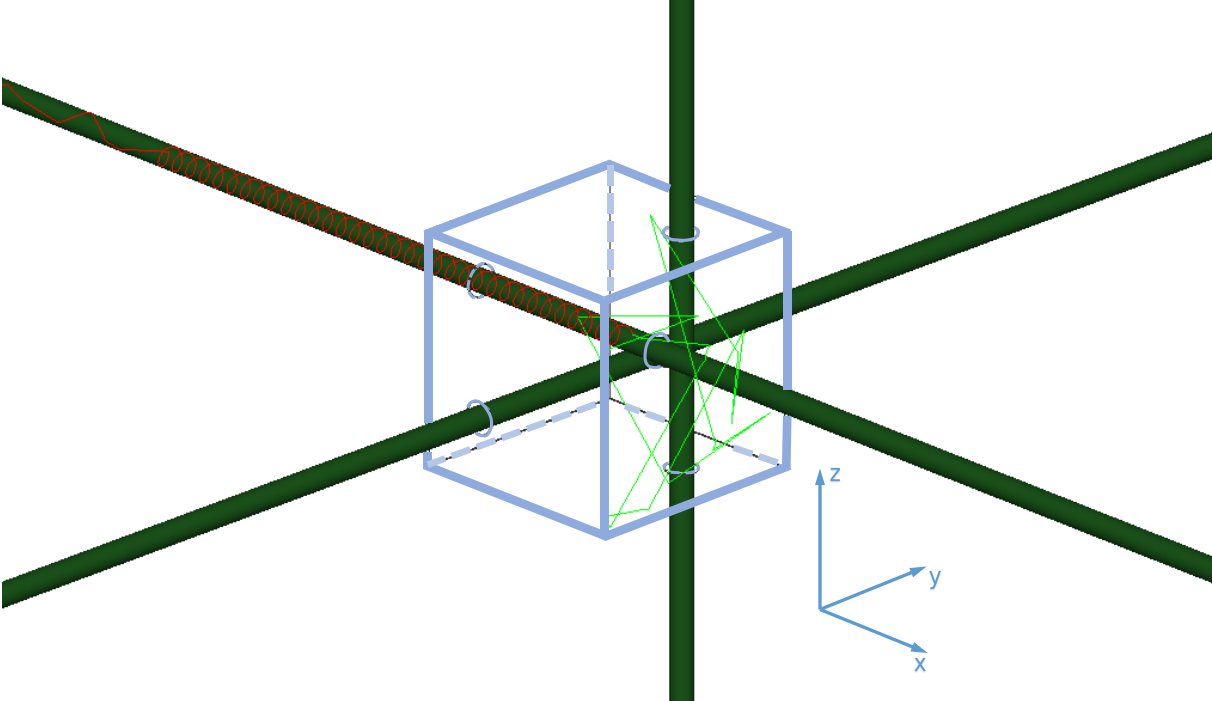}
    \caption{A simulated single SuperFGD cube with three WLS fiber insertions. An optical photon (green) was generated in the cube volume and subsequently captured by a WLS fiber after multiple reflections. The wavelength-shifted photon (red) was then propagated along the fiber until reaching the MPPC installed at one end of the fiber.}
    \label{fig:geom}

\end{figure}
As illustrated in Fig. \ref{fig:geom}, a single SuperFGD cube with three WLS fibers was depicted, showing  the travel path of an optical photon as well, in different colors depending on the undergoing optical process: its generation and multiple reflections in the optically-isolated scintillator cube (green segments), and the subsequent capture and trapping in a WLS fiber (red segments). Without loss of generality, a single 1 $\times$ 1 $\times$ 1 $\text{cm}^3$ cube with diffusing reflective boundary was simulated, being consistent with one SuperFGD unit. The diffusive nature of the boundary was simulated by setting a small value of the polishing parameter in the adopted GLISUR model for the Geant4 optical boundary \cite{AGOSTINELLI2003250}. It introduces large variations in the angle between the micro facet of the cube boundary and the macro surface normal, resulting in diffusing light. Three double-cladding Kuraray Y11 WLS fibers \cite{kuraray_y11} of 0.5 mm radius were inserted in the cube through three through-going holes of 0.75 mm radius along X, Y and Z direction respectively, each being 2.9 mm away from the corresponding cube boundary. All the fiber components were simulated, including the fiber core made of Polystylene, the inner cladding made of Polymethylmethacrylate
(PMMA) and the outer cladding made of Fluorinated polymer (FP). A complete air gap was maintained between the WLS fiber and the scintillator volume. However, it is worthy to note that, in the real detector, fibers were partly in contact with the scintillator, which may result in loss of the light propagating up through the outermost cladding. It is not trivial to simulate such effect geometrically in the simulation. Nonetheless, this effect can be incorporated into the short attenuation length of the light propagation in the fibers, thus will be integrated in the tuning of the fiber attenuation lengths. Further elaboration on this aspect, including the impact of the air gap on the light propagation, will be provided in Sec.~\ref{attenuation}. At one end of each fiber, a simple 1.3 $\times$ 1.3 $\text{mm}^2$ square volume was simulated as a photon detector. 
The PDE spectra of different type of MPPCs used in the different prototypes can be applied. 
Moreover, the distance between the WLS fiber end and the MPPC can be set in order to simulate loose coupling conditions. In this work a good agreement between data and MC has been achieved by simulating the WLS fiber end in contact with the MPPC.

\begin{figure}[h!]
    \centering

        \centering
            \includegraphics[width=0.7\linewidth]{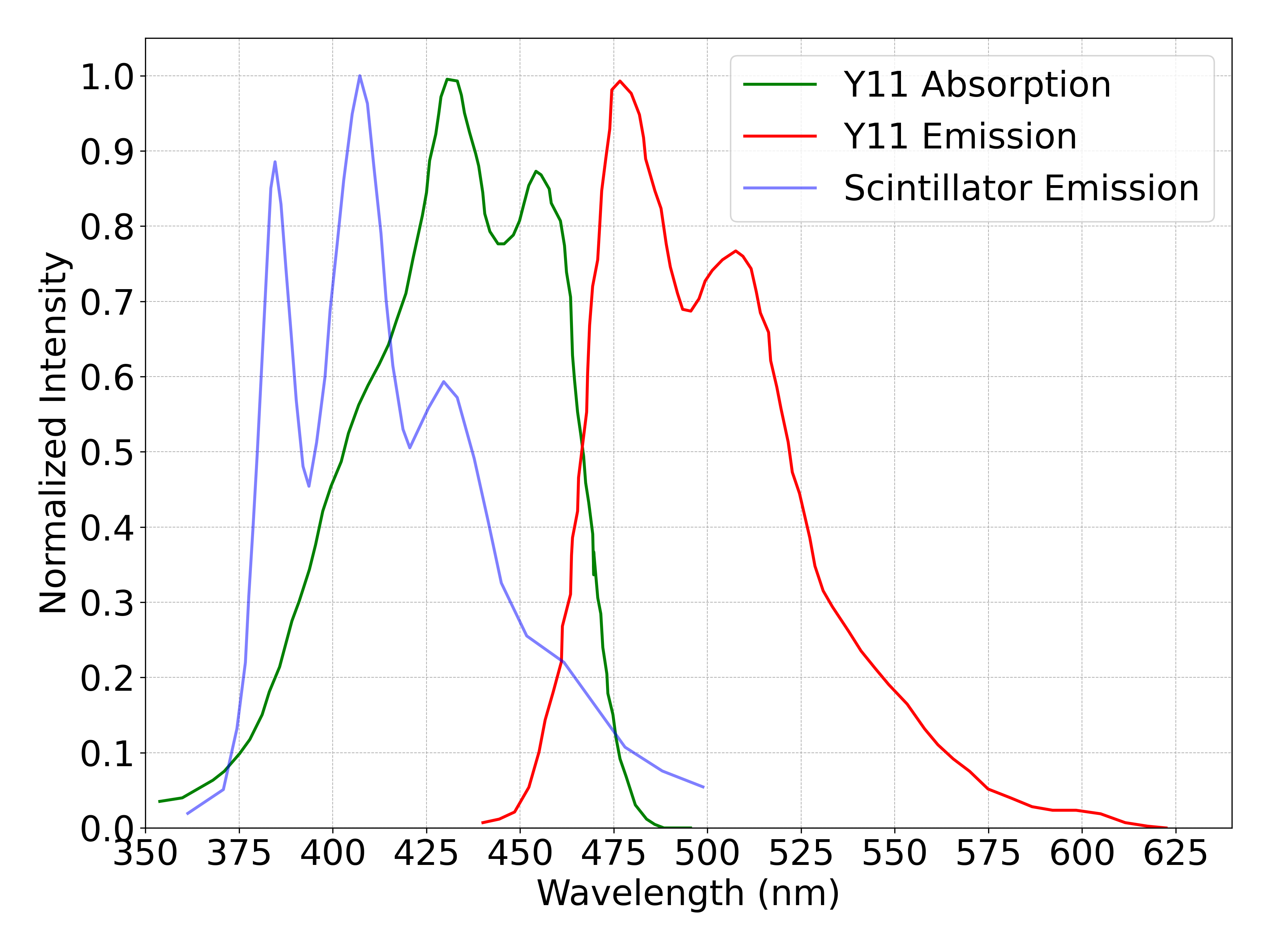}
            \caption{The implemented spectra of the plastic scintillator~\cite{POPOP} and the Y11 WLS fiber~\cite{kuraray_y11}. (Blue) Emission spectrum of the plastic scintillator. (Green) Absorption spectrum of the Y11 WLS fiber. (Red) Emission spectrum of the Y11 WLS fiber.}
               \label{fig:spec}
       
\end{figure}

A detailed simulation of the optical processes is performed for all the components of the detector unit. The Geant4 physics list "QGSP\_BERT\_HP" was adopted to accurately simulate the interacting processes of the particles with energy below 10 GeV. The optical interfaces were simulated using the "GLISUR" model \cite{AGOSTINELLI2003250}. In the scintillator volume, three major optical processes were simulated: the scintillation process controlled by the intrinsic light output, the emission spectrum and the decay time of the scintillator; the attenuation of the scintillation light in the bulk determined by an exponential function; and the light crosstalk determined by the coating reflectivity. Considering the very thin etched reflective boundary of the SuperFGD cube, zero absorption in the cube boundary was simulated. When the incident light interacts with a WLS fiber, it undergoes absorption and subsequent re-emission. The absorption and emission spectra of the WLS fiber material as well as its characteristic decay time have been provided by Kuraray \cite{kuraray_y11} and simulated. The re-emitted light within the fiber either remains confined or escapes depending on the angle of propagation, the relative refractive indices of the constituent materials of the fiber, and the surface roughness of the exterior of the fiber. In the simulation framework, the latter is modulated by a single parameter which initially assumes an ideal state of perfect polishing for the surface of the fiber. As the final step of the light propagation, when the photon reaches the MPPC, the detection is determined with respect to the implemented PDE spectrum, that for a given wavelength is treated as a binomial efficiency. To summarize, Table \ref{table:parameters} encapsulates all the pertinent optical parameters. Most of the parameters have been roughly constraint in independent measurements or provided directly by the manufacturers, and the others can vary in a relatively larger region, and are tuned on the collected detector prototype data. Figure~\ref{fig:spec} presents the adopted spectra, offering a visual representation of the key optical characteristics. 
\begin{table}[h!]
\centering
\begin{tabular}[|l|]{ |p{5cm}||p{5cm}| |p{3cm}| }
 \hline
 \multicolumn{3}{|c|}{Parameters} \\
 \hline
 Geometry& Parameter Name& Preset Value\\
 \hline\hline
 \multirow{5}{5cm}{Scintillator} & Light Yield & $\text{10 photon/keV}$\cite{pl_scint}\\ 
 \cline{2-3}
 & Attenuation Length (Cube)& 10 cm\cite{pl_scint}\\
 \cline{2-3}
 & Cube Boundary Reflectivity & 0.96\\
 \cline{2-3}
 & Emission Spectrum & \cite{POPOP}\\
  \cline{2-3}
 & Decay Time (Cube) & 1 ns\cite{pl_scint}\\
 \cline{2-3}
 & Refractive Index  & 1.59\cite{pl_scint}\\
 \hline\hline
 \multirow{6}{5cm}{WLS Fiber} & Outer Surface Roughness & 0.98\\
 \cline{2-3}
 & Absorption Spectrum & \cite{kuraray_y11}\\
 \cline{2-3}
 & Emission Spectrom & \cite{kuraray_y11}\\
 \cline{2-3}
 & Decay Time (Fiber) & 8 ns\cite{Fiber_decay_Brekhovskikh:2000pna}\\
 \cline{2-3}
 & Attenuation length (Fiber)& 4.2 m \cite{kuraray_y11}\\ 
 \cline{2-3}
 & Refractive Index (Core) & 1.59\cite{kuraray_y11}\\
 \cline{2-3}
 & Refractive Index (Inner Clad) & 1.42\cite{kuraray_y11}\\
 \cline{2-3}
 & Refractive Index (Outer Clad) & 1.49\cite{kuraray_y11}\\
 \cline{2-3}
 & Painting reflectivity* & 0.8\\
 \hline\hline
 MPPC & Photon Detection Efficiency (P.D.E.) vs wavelength & \cite{hamamatsu_s13360-1325cs}\\
 \hline\hline
 \multicolumn{3}{|p{13cm}|}{*The painting at one end of the fiber is only implemented specifically in the simulation of the 2018 prototype, and the fiber attenuation test where the painting was applied.} \\
 \hline

\end{tabular}\
\caption{Parameters corresponding to each geometry component. Most of the parameters with a referenced publication have been fixed in the simulation, while the ones without, as well as a few parameters only roughly provided by the reference, have been tuned on the data collected with the different prototypes as described in Sec.~\ref{sec:dataset}.}
\label{table:parameters}
\end{table}

\subsection{Tuned parameters of the optical model}
\label{sec:optsim;4}
Taking advantage of multiple independent measurements about the scintillator material and the WLS fibers, many of the simulation parameters in Tab.~\ref{table:parameters} can be inferred and constrained to a relatively accurate level. However, certain parameters, in particular the cube boundary reflectivity, fiber surface roughness, and the scintillator attenuation length, require further tuning using the data collected with the SuperFGD prototypes, either from the cosmic muons or the beam tests, as described in Sec.~\ref{sec:dataset}. In this section, we focus on the model tuning and validation based on several major optical observables that characterize the detector performance, including the cube light yield per channel per MIP, the cube-to-cube crosstalk, the WLS fiber attenuation length, the cube light yield non-uniformities and the detector time resolution. Unless specified otherwise, the simulations involved mono-energetic muons of 500 MeV, to reproduce MIPs from beam tests and cosmic particles. In this scenario, we anticipated minor difference between the simulation and the experimental data coming from the less than 5\% difference in the energy deposition in a carbon target.

\subsubsection{Tuning of the light response}\label{sec:LY}
In this section the beam test results
described in Sec.~\ref{sec:2018_data}
were taken as a reference. The WLS fiber length was 1.3 m with about 4 cm protruding at one side out of the prototype. One end of the fibers was covered with reflective aluminized-based paint, while the other end coupled to Hamamatsu MPPCs of type S12571-025C of 25 \textmu m pitch size~\cite{hamamatsu_s13360-1325cs}.

The light yield of a single fiber channel is determined by a series of optical processes, spanning from the initial generation of scintillation light to the final readout. Key parameters contributing to the processes include: 1) the light output of the plastic scintillator; 2) the effective attenuation length of the polysterene-based plastic scintillator, which describes effectively both the light loss in the scintillator cube volume and in the reflective coating, where the light loss might be caused by the multiple internal reflections in the coating volume, potentially not negligible even if the thickness is small; 3) the reflectivity of the white coating layer, which determines the light leakage from a cube to another, while the light absorption of the coating was assumed to be negligible; 4) the absorption length of the WLS fibers; 5) the refractive index of the fiber core and the two claddings; 6) the roughness of the fiber outer cladding; 7) the reflectivity of the aluminized layer which is specifically used for such prototype; 8) the photon detection efficiency of the MPPC.

The tuned parameters shown in Tab.~\ref{table:parameters} are the cube boundary, the fiber surface roughness, and the reflectivity of the aluminized paint reflectivity.
An initial value of 0.96 was set to the cube boundary reflectivity, compatible with typical reflectivity values used in similar contexts and materials \cite{3DET:2022dkw}. Such parameter value was validated by comparing the simulated light crosstalk with the dataset described in Sec.~\ref{Xtalk}. Negligible absorption in the etched cube boundary was assumed, rationalized by its minimal thickness of 0.05 - 0.08 mm. The roughness of the fiber outermost surface, being vulnerable to the environmental damage, influences the light yield by controlling the amount of light trapped within the total internal reflection cone determined by the refractive index between the cladding material (n=1.42) and the air (n=1). This is often considered to be responsible of the short attenuation length of optical fibers, usually well below 1 m. The inner surfaces, e.g. the surface between the fiber core and the inner cladding, on the other hand, are always considered perfect since they were not directly exposed to environment. A perfect polished fiber outer surface was assumed as the initial setup. Then, the parameter was further tuned and cross-validated by both the light yield test and fiber attenuation curve test in Sec.~\ref{attenuation}. The last free parameter is the effective reflectivity of the aluminized painting at the fiber end. The optimal value was found by scanning the parameter space within [0.76, 0.9], fitting the simulated light yield and the measured data. Among the three free parameters, the coating reflectivity can be roughly determined independently from the crosstalk measurement with the beam test data. Several optical simulations were performed with different sets of values of the non-constrained parameters and the one that reproduced the best all the available datasets was chosen as the final combination.

The simulation chain consists a few different steps.
A muon with a kinetic energy of 500 MeV is shot along the z-axis with a starting point randomly sampled from a uniform distribution on an x-y plane (front vertical) 5 cm away from the cube volume. As the simulated muon traverses the cube, it deposits an average energy of 1.8 MeV, following Gaussian-Landau distribution, leading to the generation of scintillation photons following approximately the Poisson statistic with a mean light output of 10k photons per MeV. Statistically, of these photons, 53\% were found to be absorbed in the scintillator, 17\% escaped as crosstalk and the remaining 30\% were captured and wavelength-shifted by three WLS fibers, following again a Poisson statistic. The wavelength-shifted photons are emitted isotropically. Some of the photons are geometrically trapped depending on their incident angles and positions. Given perfect outer surface and that fibers are not in contact with the scintillating volume, about 30\% of the re-emitted light, including both meridional light and skew light, is internally reflected within the light cone determined by the refractive indices of the fiber materials \cite{trap_Achenbach:2004gi}, among which 10\% remains in the core and 20\% is trapped up to the outermost cladding. The later is significantly influenced by the surface condition of the fiber, which smears the reflected angle of the light. After about 1 m propagating within the fiber, about 80\% of the trapped photons survive and reach the MPPC, where about 35\% photoelectrons (P.E.) can be detected. 

\begin{figure}[h!]
    \centering
    \includegraphics[scale=0.6]{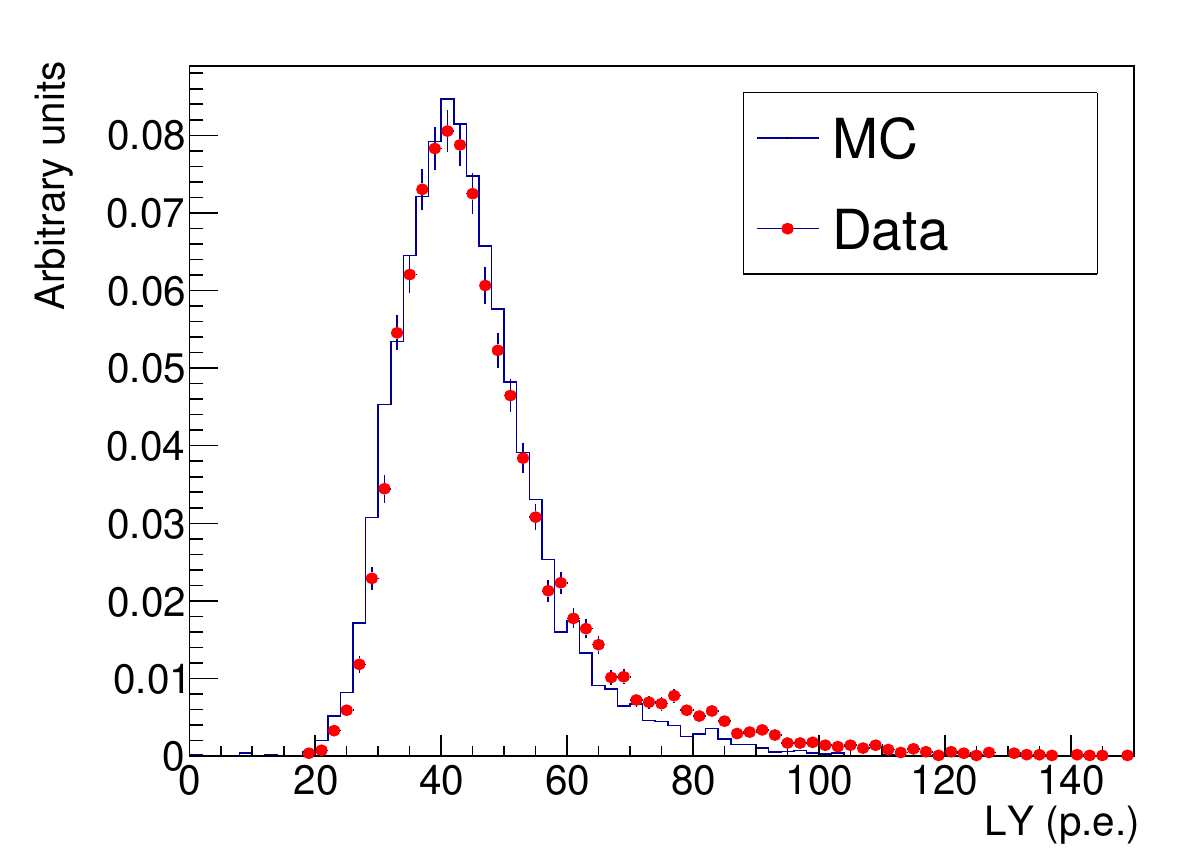}
        \caption{(Blue histogram) The simulated one fiber light yield distribution of 500 MeV muons propagating along z direction, randomly generated within x-y plane. (Red dots) the measured data of the 5 x 5 x 5 SuperFGD prototype \cite{sfgd-2018}.
        \label{fig:LY_MC}
}
\end{figure}

An optimal simulation model was found with the parameters shown in Tab.~\ref{table:parameters}. Figure~\ref{fig:LY_MC} displays the simulated light yield distribution compared to the measured data. The data and MC distributions are in good agreement.

\subsubsection{Light crosstalk}\label{Xtalk}
According to the measured data, the average crosstalk probability is around 3-3.5\% per side,
slightly varying between the different measurements 
\cite{sfgd-2018, sfgd-Blondel:2017orl, Blondel:2020sfgd, T2K:2019tech}. In this section we use the beam test result of the larger SuperFGD prototype made of 24 $\times$ 8 $\times$ 48 cubes, where an average crosstalk of 2.9\% per face was more accurately derived with highly-ionizing proton events \cite{Blondel:2020sfgd}. The crosstalk is typically calculated as the ratio between the light yield of parallel channels in the main hit cube and a neighbour cube respectively. In order to optimize the simulation speed while keeping the maximum model flexibility, only a single cube was simulated. In such a way, it enables us to study the intrinsic light leakage using the tuned model which is not accessible in the beam tests. On the other hand, to compare with the measured data, the intrinsic light leakage was converted into the form of light yield by introducing auxiliary random processes based on the modeled light propagation processes from the cube to the MPPC through the WLS fibers. The details are described in this section. 

We approached the crosstalk fitting through a two-steps process. In the first step, we obtain the intrinsic light leakage ratio, i.e., the ratio between the total number of photons escaping the cube and the total number of photons generated. The average light leakage ratio is statistically consistent with the average crosstalk. In this step, the simulation parameters were tuned in the first order. In the second step the intrinsic light leakage was converted into the light yield with the help of the later introduced random sampling analysis procedure, and compared to the measured crosstalk distribution, to validate the tuned model.

As discussed above, we focused on the light leakage ratio in the first step. Being independent from the propagation within the WLS fiber, it was determined by three different correlated processes in the cube volume: the light leakage through the cube faces, the light attenuation within the scintillator volume, and the light capture by the WLS fibers. The later two processes are determined by the cube attenuation length and the WLS fiber absorption spectrum respectively, which are pre-fixed by external data (Tab.~\ref{table:parameters}), while the light leakage, fully controlled by the cube boundary reflectivity, is to be tuned. Simulations with different values of the cube boundary reflectivity were run to find the best fit of the average measured crosstalk. As a result, the 18\% total light leakage through six faces was obtained with a reflectivity of 0.96, being consistent with the measured crosstalk of 2.9\% through one reflecting face. The cube boundary reflectivity was then fixed in the model fitting. 

\begin{figure}[h!]
    \centering
    \includegraphics[width=0.66\linewidth]{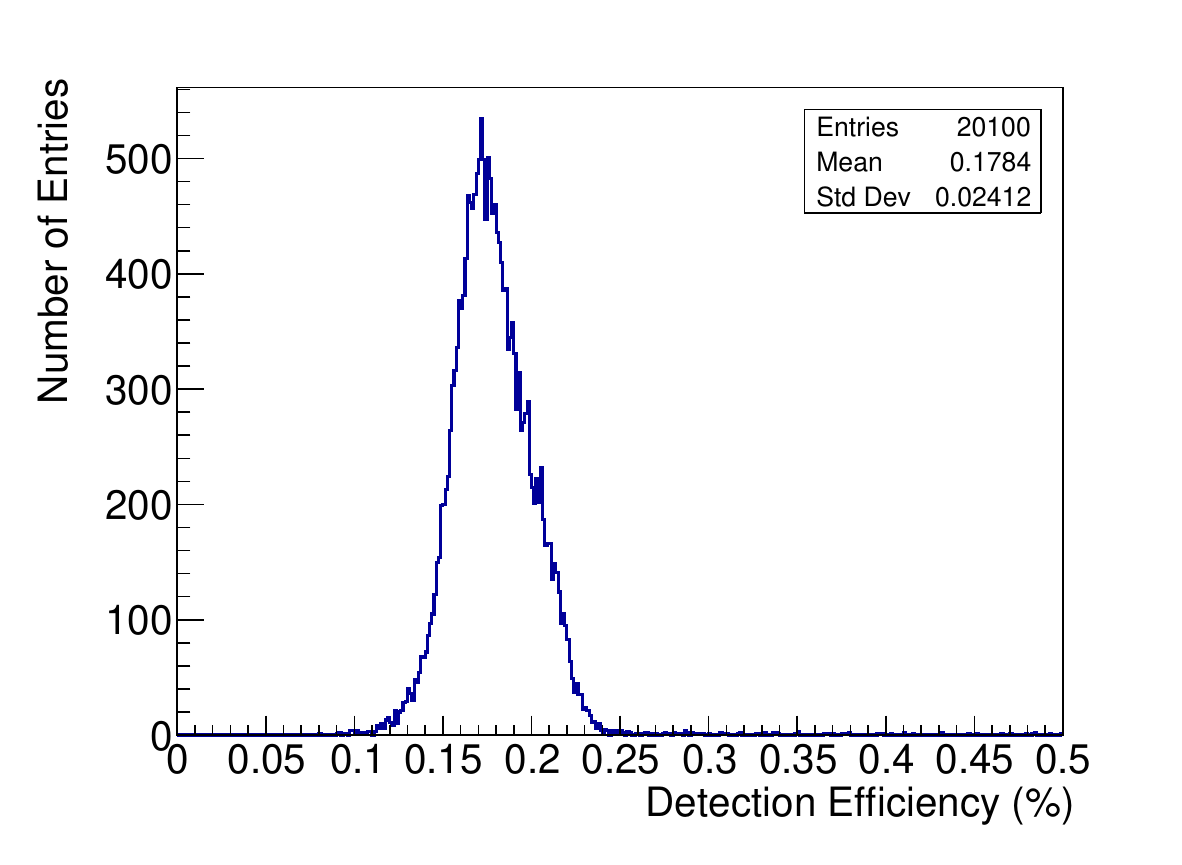}
        \caption{Ratio between the one fiber light yield and the number of primary scintillation photons generated in the scintillator volume. It represents the distribution of the effective detection efficiency of a photon generated in a single SuperFGD cube and detected by a single channel. 
        }
        \label{fig:LighttoLY}
\end{figure}

\begin{figure}[h!]
    \centering
    \includegraphics[width=0.63\linewidth]{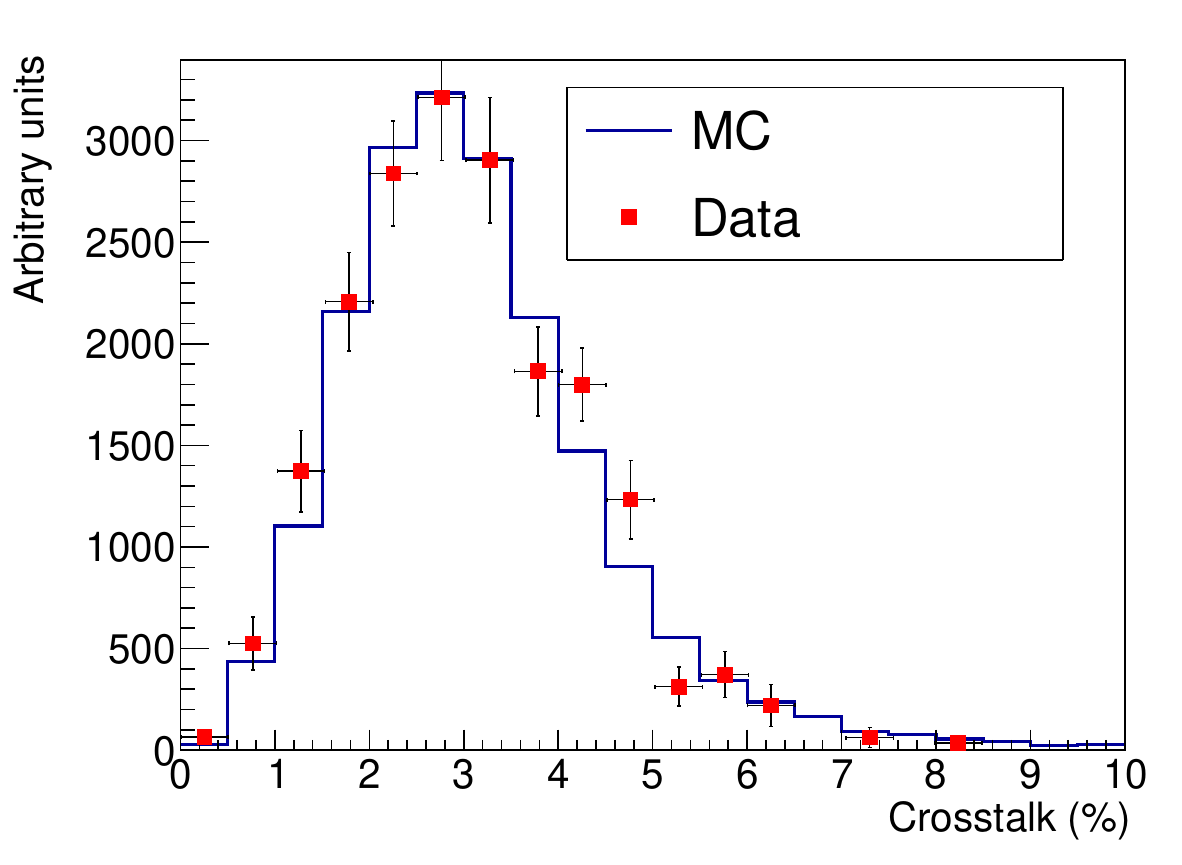}
        \caption{Distribution of the optical crosstalk for a single cube
        %, in \%, 
        obtained from the optical simulation of 500 MeV muons propagating along z direction, randomly generated within the x-y plane. The red data points were digitized from the beam test data described in \cite{Blondel:2020sfgd}. The blue histogram is obtained from the optical simulation.
        }
        \label{fig:Xtalk_MC}
\end{figure}

After the model parameters were fixed (details can be found in Sec.~\ref{sec:LY} and Sec.~\ref{attenuation}), we extracted the statistical distribution of the cube-to-cube cross talk adopting a random sampling procedure and compare it with the one from the beam test data.
We estimated the light leakage through a single reflecting face by statistically sampling the generated primary photons, employing a binomial distribution with a 18\% leakage probability 
obtained from the previous analysis step. Then, each of the escaped photons was randomly sampled from integers between [1,6] to determine from which face. 
An auxiliary distribution of the ratio between the cube light yield and the initial generated number of photons was obtained from the optical simulation as in Sec.~\ref{sec:LY}. 
Such distribution represents the effective photon detection efficiency in a single SuperFGD cube with mean value of 0.18\%, as shown in Fig.~\ref{fig:LighttoLY}. 
For each leaked photon sampled, the detection efficiency was randomly extracted from this distribution and the crosstalk was obtained by collecting all the photon detected.
Finally, the simulated crosstalk was calculated by taking the ratio between the number of the detected photons in the readout channel of the neighbouring cube and the simulated light yield from the channel in the main cube channel, crossed by the charged particle. 
The result is shown in Fig.~\ref{fig:Xtalk_MC}. 
The data crosstalk distribution is quite well reproduced by the optical simulation.  

\newpage

\subsubsection{WLS fiber attenuation length} \label{attenuation}
The optical photons captured by the fibers undergo again three primary optical processes: wavelength shifting, light trapping, and light attenuation. The wavelength shifting process is governed by the fiber absorption and emission spectra, while the light trapping depends on the refractive index of the fiber material. The light attenuation is more complex, influenced by several factors. It can be empirically described by the following function:

\begin{equation}
\begin{split}
    LY(x) =& LY_0 \left[ \alpha ~ \exp{(-\frac{x}{A_L})} + (1-\alpha)\exp{(-\frac{x}{A_S})} \right]\\ 
    +& R ~ LY_0 \left[ \alpha ~ \exp{(-\frac{2L-x}{A_L})} + (1-\alpha)\exp{(-\frac{2L-x}{A_S})} \right].
  \label{eq:attenuation}
\end{split}
\end{equation}

Here $LY(x)$ represents the light intensity after traveling a distance $x$ in the WLS fiber. The first term describes the direct light propagating towards the MPPC and is characterised by a double exponential function. $LY_0$ is the overall light yield normalization, $\alpha$ is the ratio of the light component of long attenuation length, while $A_L$ and $A_S$ are, respectively, the long and the short attenuation lengths. The second term describes the reflected light from the uninstrumented end of the WLS fibers. Its reflection factor is $R$. The same double exponential function is used but, this time, with a propagation length of $2L-x$, where L is the total length of the WLS fiber. The long attenuation length ($A_L$) describes the stable component of light, also known as meridional light, trapped within the WLS fiber core, while the short attenuation length ($A_S$) accounts for the light that is reabsorbed again by the same WLS as well as cladding or skew light. The cladding light can be lost due to factors like dust, surface defects or contact with external materials. The skew light can escape due to a slight bending of the WLS fiber. While the manufacturer specifies $A_L$ as more than 3.5 meters \cite{kuraray_y11},
$A_S$ is harder to model properly and requires further fine-tuning.
In the simulation, the light propagation is simulated using the implemented optical model of Geant4 software, which considers every physical process, rather than relying on Eq.~\ref{eq:attenuation}. The absorption length of the re-emitted photons, which governs the long attenuation length component, was set to 5.5 m in the simulation. The number was chosen to replicate the technical attenuation length fitted with Eq.~\ref{eq:attenuation} on data, i.e., about 4.2 m.

\begin{figure}[h!]
    \centering
    \includegraphics[width=0.48\linewidth]{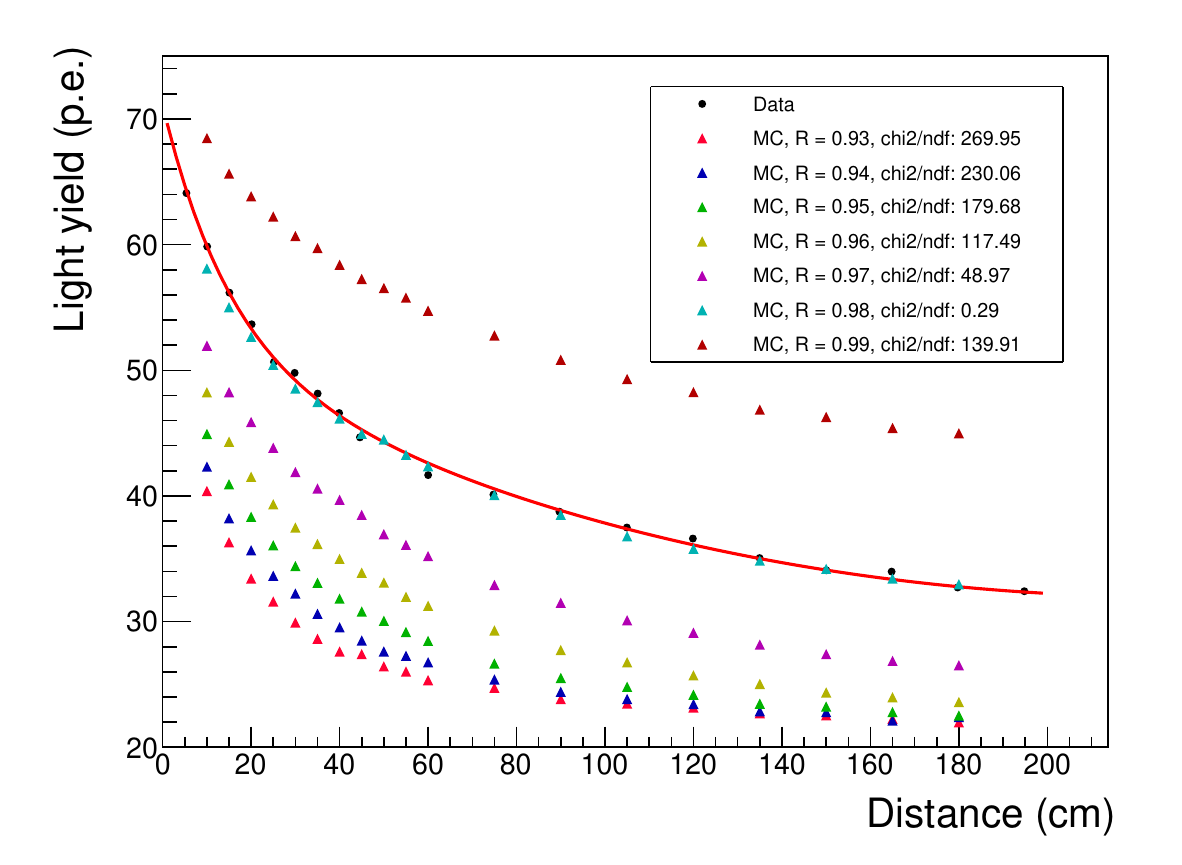}
        \includegraphics[width=0.48\linewidth]{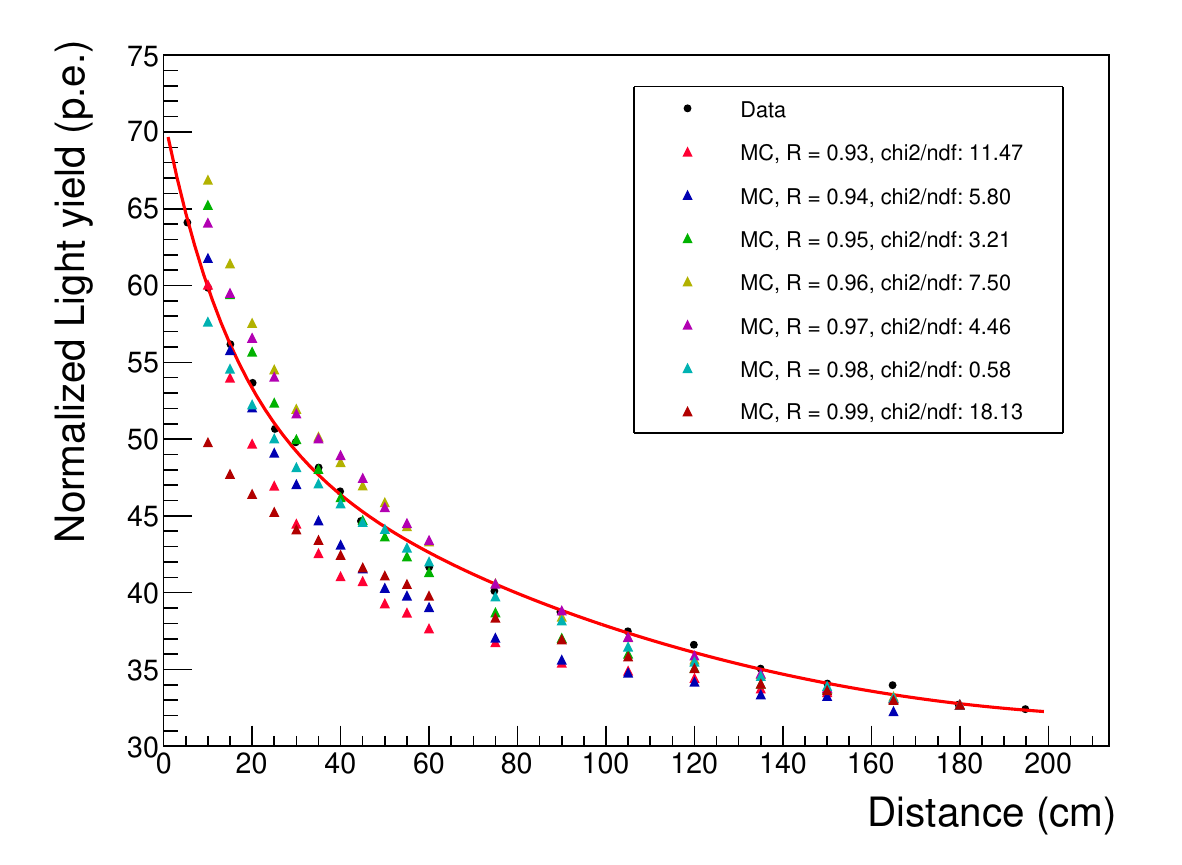}
    \caption{MC attenuation length of the WLS fibers. Each color corresponds to a different value of the outer surface roughness (R). Black dots correspond to the measured data. The $\chi^2$ value, result of the data-MC comparison, is shown for different roughness values.
    While left figure provides a data-MC comparison of both the shape and the normalisation of the attenuation length, all the curves in the right figure were rescaled to equalise the light yield at the distance of 180 cm to obtain an unambiguous analysis of the shape.}
    \label{fig:attcurve_MC}
\end{figure}

The short attenuation length was simulated by accounting for the overlap in the absorption and emission spectra as well as for the light loss through the cladding surface. While the former one is governed by the provided spectra, the latter one is mainly controlled by the surface roughness of the WLS fiber, a factor defined in Geant4 by a single parameter that dictates the norm fluctuation of micro facet relative to the surface plane. Significant difference in light yield has only been observed when the surface is close to a perfect condition, while the light yield is almost unchanged when the surface roughness is smaller than 0.9 since most of the cladding light was already lost before reaching the MPPC under this condition. Thus, different values from 0.93 to 0.99 were scanned to find the optimized configuration. At 0.93, the surface roughness is such that a relatively large amount of the light reaching this surface escapes quickly after a few internal reflections, whereas at 0.99, representing an almost perfectly polished clean surface, the light is mostly trapped within the fiber, as long as the total internal reflection condition is satisfied. 
The measurement using a 2 m long WLS fiber with one end covered by reflective paint as shown by the blue curve in Fig.~\ref{fig:fiber_attenuation} was taken as the reference. This choice was dictated by the fact that the same configuration was used in the reference measurement for the light yield tuning in Sec.~\ref{sec:LY}.
Figure.~\ref{fig:attcurve_MC} shows the comparison between data and MC. Overall, the data-MC best fit is given by the outer surface roughness of 0.98.

\subsubsection{Light yield spatial non-uniformity}

In this section the non-uniformity of the light yield as a function of the particle incoming position in the cubic scintillator volume is discussed. Theoretically, this characteristic is inherently linked to the attenuation length of the chosen plastic scintillator.
On the other hand, it is also influenced by the reflectivity and the absorbance of the white reflector, that acts as an additional effective light attenuator. A shorter attenuation length typically results in a stronger correlation between the interaction point and the measured light yield. 
Moreover, The light generated by a particle interacting near a WLS fiber is more likely captured by the fiber, leading to a higher light yield. 
Conversely, longer attenuation lengths would result in a more homogenous distribution of the light yield, regardless of the spatial location of the interaction point. 
Understanding this variance across the cube volume is crucial for the model tuning, particularly in validating the scintillator preset attenuation length, which, in the current setup, is established at the level of 10 cm \cite{pl_scint}, as highlighted in Tab.~\ref{table:parameters}. 
The reference data used for this analysis are described in Sec.~\ref{sec:tohoku-beam-test}.

\begin{figure}[h!]
    \centering
    \includegraphics[width=0.7\linewidth]{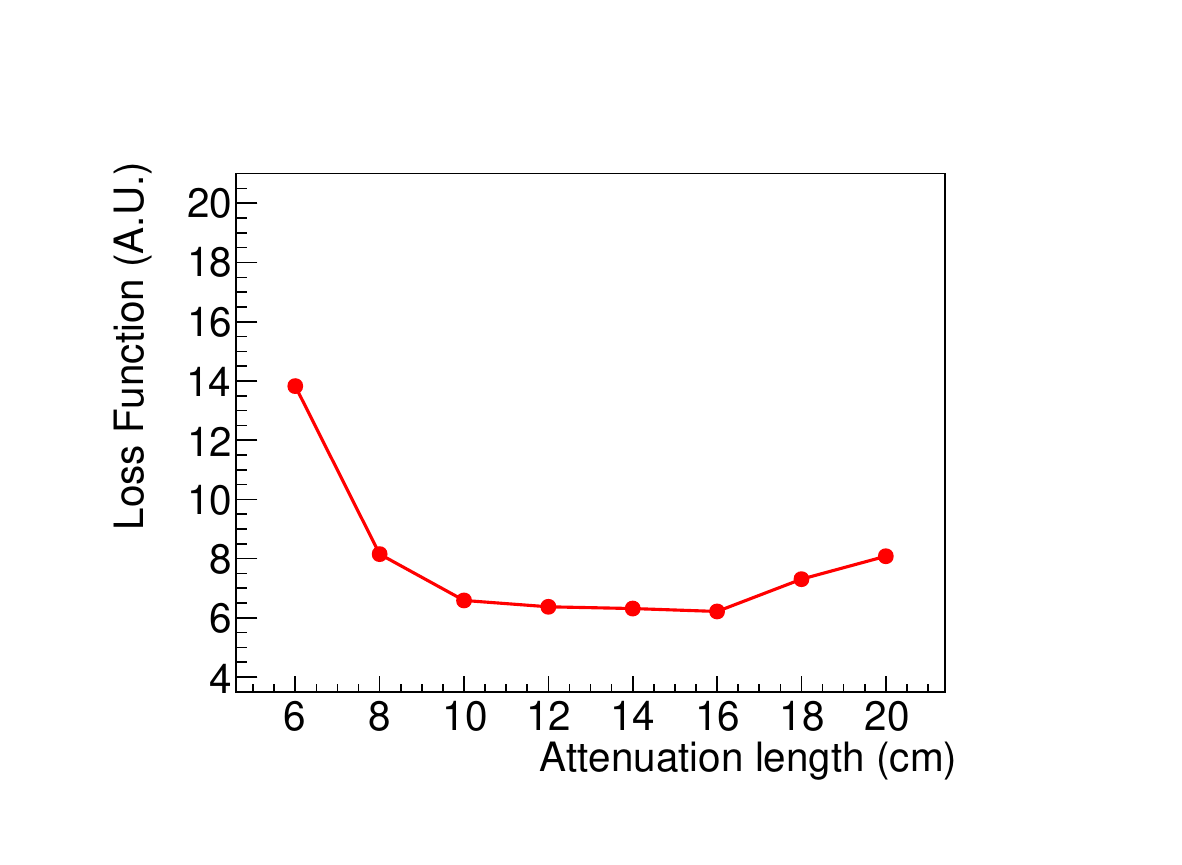}
    \caption{The calculated total loss function of the light yield maps of the three readout fiber channels between the data (see Sec.~\ref{sec:tohoku-beam-test}) and the simulation as a function of the scintillator attenuation length.}
    \label{fig:nonuni_chi2}
\end{figure}

\begin{figure}[!h]
    \centering

        \begin{subfigure}{0.3\textwidth}
            \includegraphics[width=\linewidth]{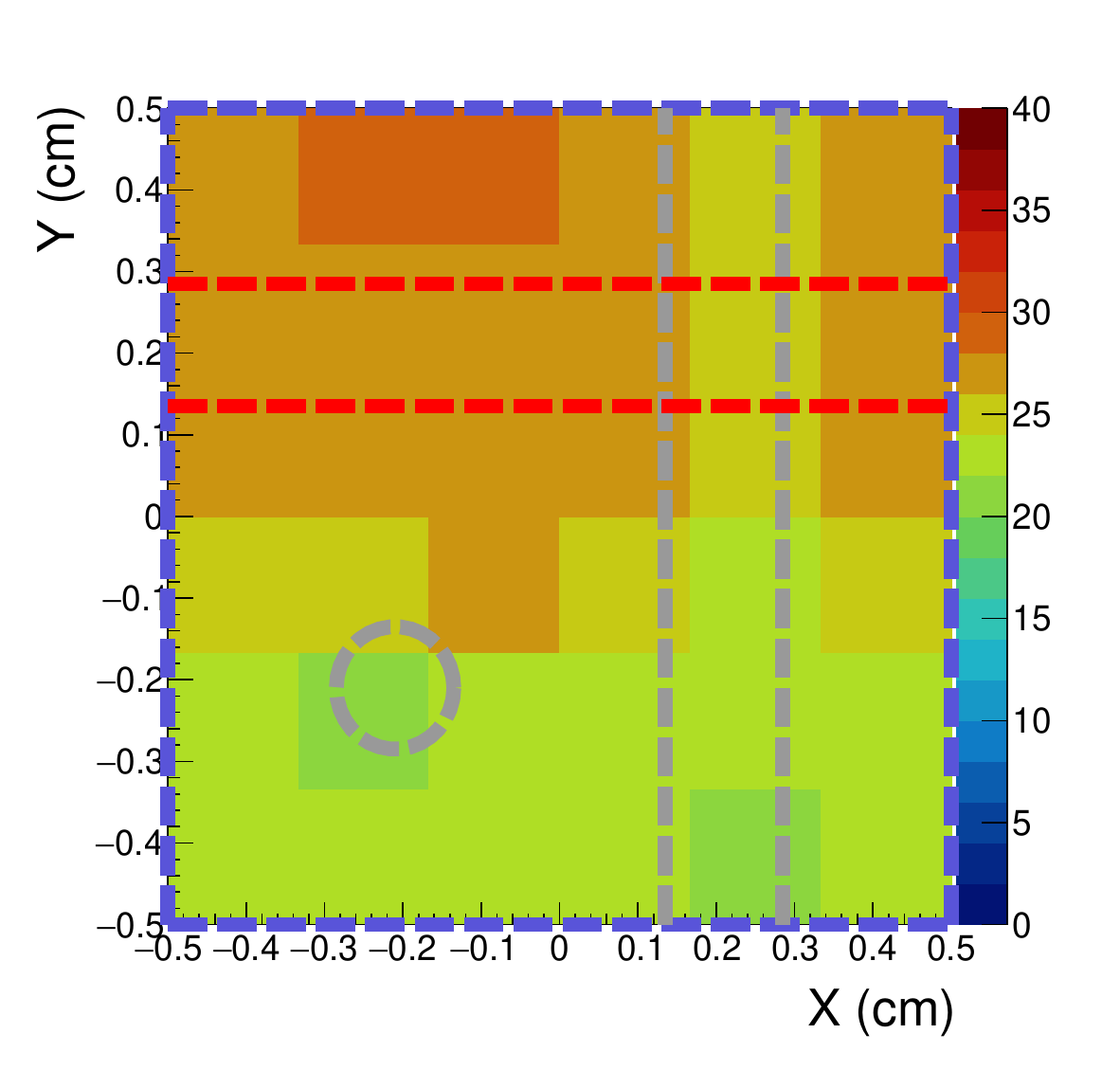}
            \caption{X Fiber readout}
        \end{subfigure}%
        \begin{subfigure}{0.3\textwidth}
            \includegraphics[width=\linewidth]{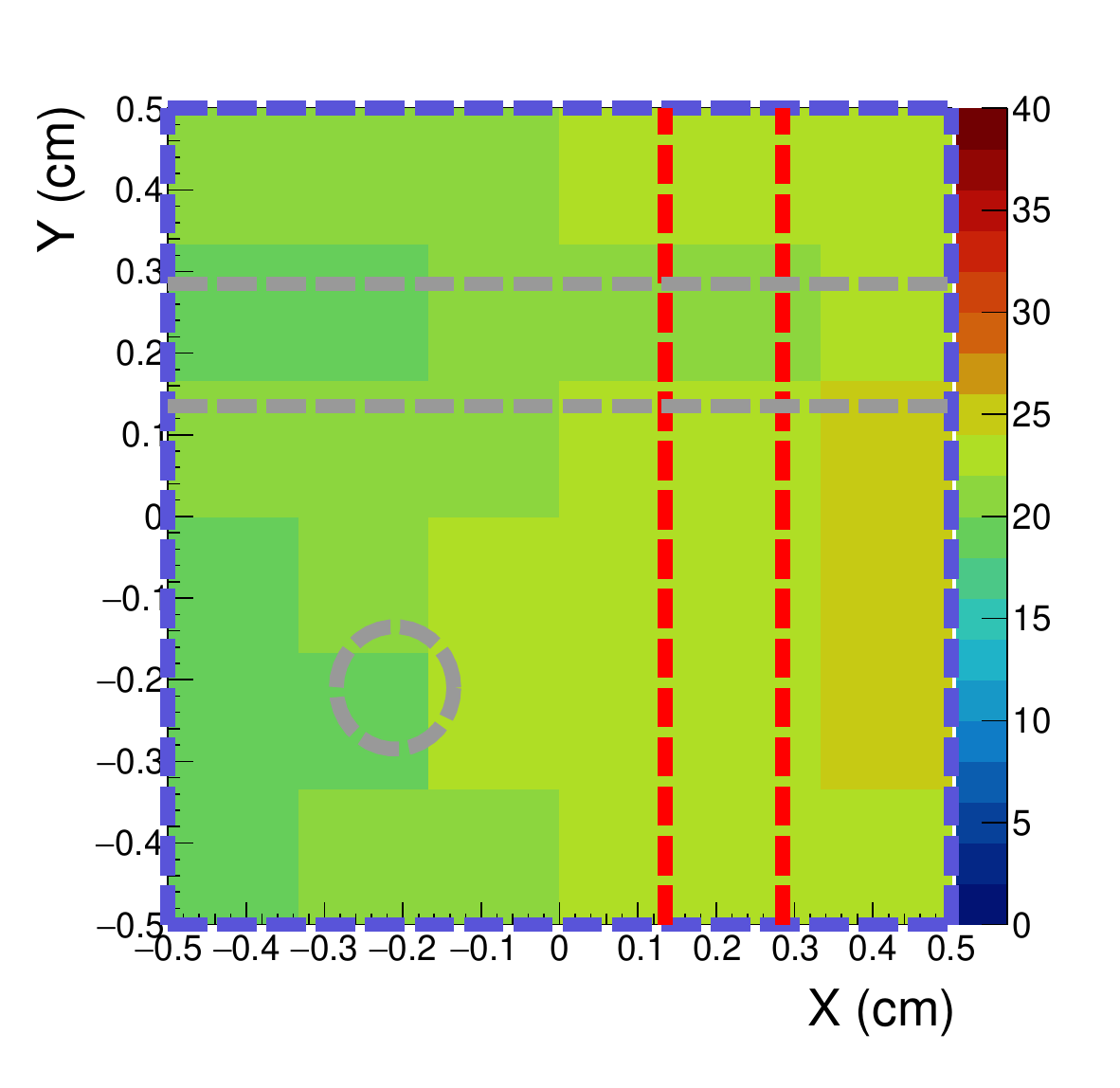}
            \caption{Y Fiber readout}
        \end{subfigure}
        \begin{subfigure}{0.3\textwidth}
            \includegraphics[width=\linewidth]{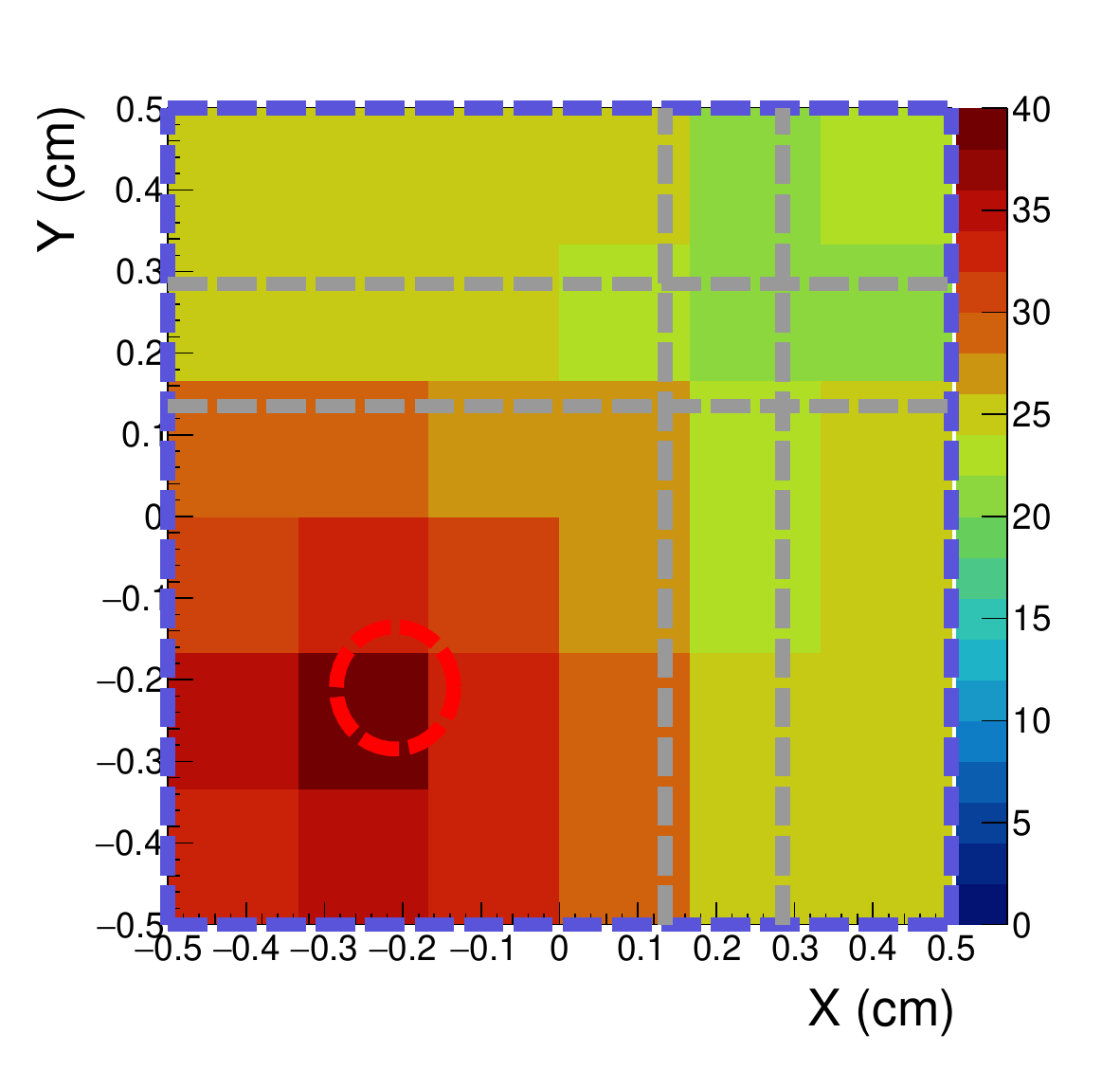}
            \caption{Z Fiber readout}
        \end{subfigure}
        \medskip
        \begin{subfigure}{0.3\textwidth}
            \includegraphics[width=\linewidth]{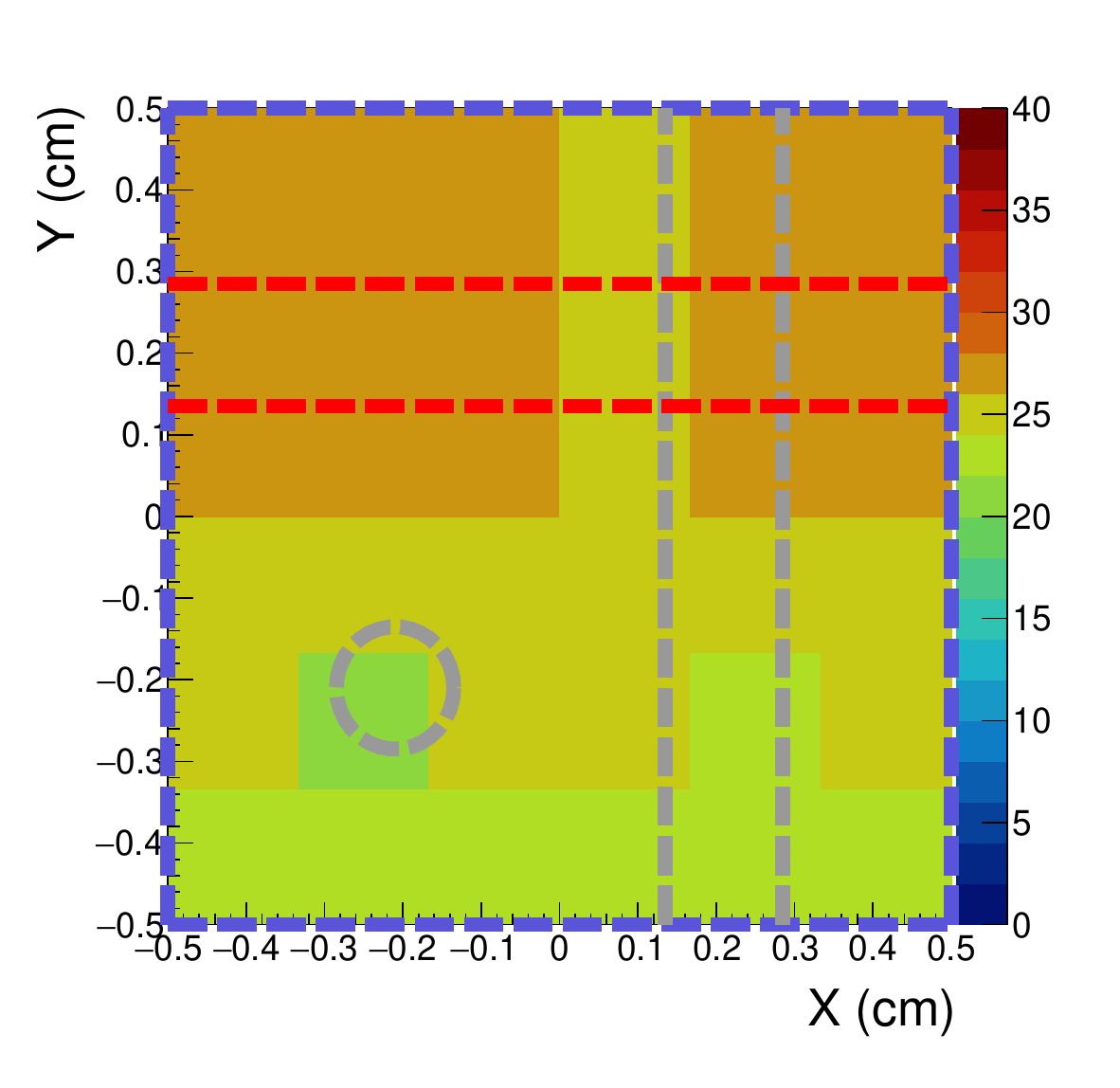}
            \caption{X Fiber a readout}
        \end{subfigure}%
        \begin{subfigure}{0.3\textwidth}
            \includegraphics[width=\linewidth]{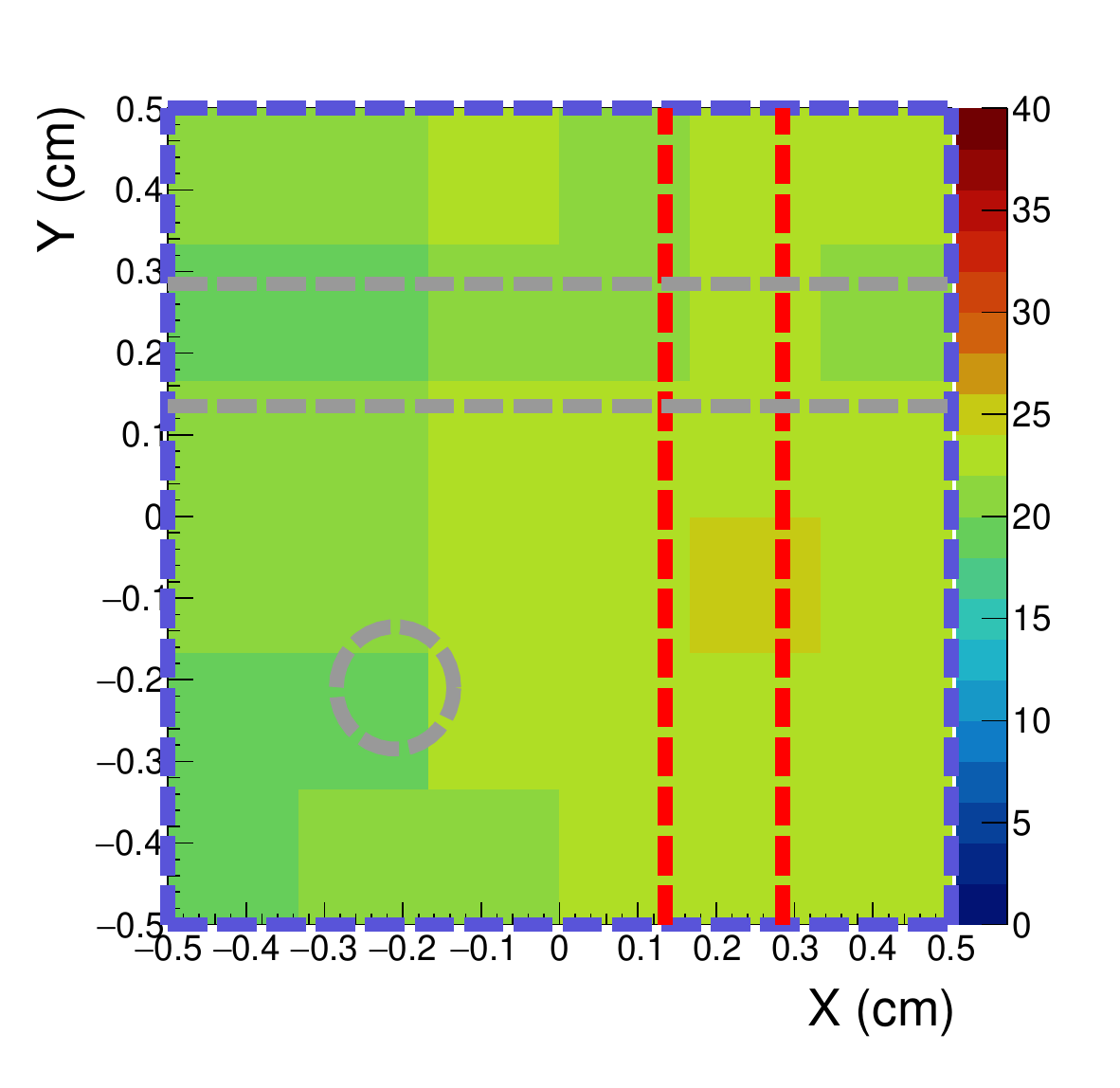}
            \caption{Y Fiber readout}
        \end{subfigure}
        \begin{subfigure}{0.3\textwidth}
            \includegraphics[width=\linewidth]{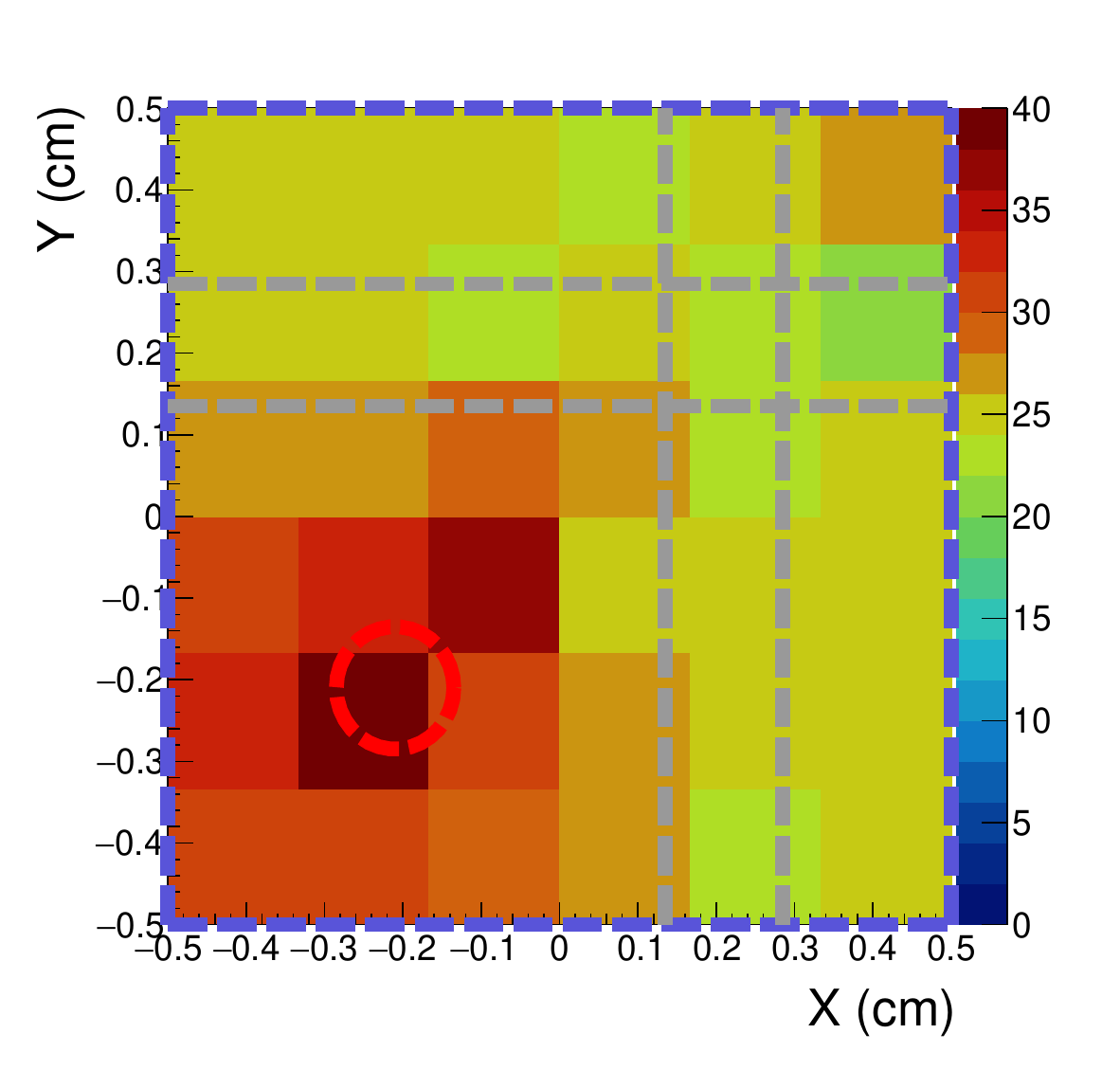}
            \caption{Z Fiber readout}
        \end{subfigure}
        \caption{Light yield uniformity map in MC (top) and data (bottom). 
        Top: simulated light yield (p.e.) detected by the MPPCs as a function of the positron position along the
        X, Y and Z axes. The scintillator attenuation length was set to be 10 cm and 500 MeV 
        positrons were simulated. 
        Bottom: measured light yield (p.e.) at the Tohoku beam test (see Sec.~\ref{sec:tohoku-beam-test}) 
        detected by the MPPCs as a function of the positron position along the X, Y and Z axes.
        In both cases, the positron beam was along the z axis. The cube cross section was shown by blue dashed lines, while the position of the holes were shown in grey. The measuring fiber channel of each map was highlighted in red.
        }
        \label{fig:nonuni_MC}

\end{figure}

In the simulation, 500 MeV positrons were uniformly shot into a single SuperFGD cube along the z axis, being consistent with the beam test setup. The starting points of the positrons were distributed uniformly on the x-y plane. Different attenuation lengths ranging from 6 cm to 20 cm were simulated in step of 2 cm.
The light yield non-uniformity maps were produced for both the measured data and the simulation. The xy cross section of a single cube (1 $\times$ 1 $\text{cm}^2$) was split into 6 $\times$ 6 square pixels. For each data set, three maps were produced showing respectively the position dependent light yield of x, y, and z fiber channels, while each pixel shows the average light yield of the fiber channel of the positrons traversing the corresponding position of the cube. Figure~\ref{fig:nonuni_MC} shows the best fitting of the non-uniformity maps compared with the same map produced from the measured data. To quantize the similarity of the non-uniformity maps between the data and simulation, a loss function was defined as follow:
\begin{equation}
f(\lambda) = \sum_{x,y,z}\sum_{i\in{\text{all pixels}}}\frac{(LY_{MC}(i,\lambda)-LY_D(i))^2}{LY_D(i)}.
\end{equation}
Here $\lambda$ represents the simulated attenuation length of the scintillator volume. The overall loss function was calculated by summing up the relative squared deviation between the measured and simulated light yields in all the pixels in all three readout maps.
The result of the loss function with respect to different scintillator attenuation lengths is shown in Fig.~\ref{fig:nonuni_chi2}. 
The 10~cm attenuation length, provided by external measurements (see Tab.~\ref{table:parameters}), falls within the flat best fit region that extends up to 16~cm.
Hence, the attenuation length of 10 cm has been fixed and kept in the final optical simulation. 

Applying the parameter producing the best fit, The light yield uniformity map is shown in Fig.~\ref{fig:nonuni_MC} and compared to the data collected at the Tohoku beam tests (see Sec.~\ref{sec:tohoku-beam-test}). %\bl{
The cross section of a single cube was represented by the blue dashed lines, while the grey dashed lines shows the position of the holes for the WLS fibers. The corresponding readout fiber channel of each map was highlighted in red.
The light yield uniformity in data and MC show a very similar pattern. 
As expected, a higher light yield is found for those particles interacting in the proximity of the WLS fibers as well as for the WLS fiber along the z axis, which is the direction of the beam positrons. 

\subsubsection{Time response and non-uniformity}

One of the key features of the SuperFGD is its sub-ns time resolution \cite{TR_Alekseev:2022jki}, affected by two processes: 
the scintillation decay time of the doping components (1.5\% of paraterphenyl (PTP) and 0.01\% of 1,4-bis benzene (POPOP)) in the polystyrene matrix; the WLS fiber wavelength shifting process, determined by the fiber doping. 
The simulated scintillation process adopted the empirical decay parameter of 1 ns, predominantly influenced by the second dopant (POPOP) \cite{pl_scint}. The decay time of the WLS fiber, on the other hand, can be found with slightly different values in literature \cite{Fiber_decay_Alekseev_2022, Fiber_decay_Brekhovskikh:2000pna, Fiber_decay_Mineev:2011xp}. 
A pre-set value of 8 ns was chosen as a compromise, being consistent with the value provided by the manufacturer \cite{kuraray_y11}. The other parameters are fixed as shown in Tab. 1.

\begin{figure}[h!]
    \centering
    \includegraphics[width=0.6\linewidth]{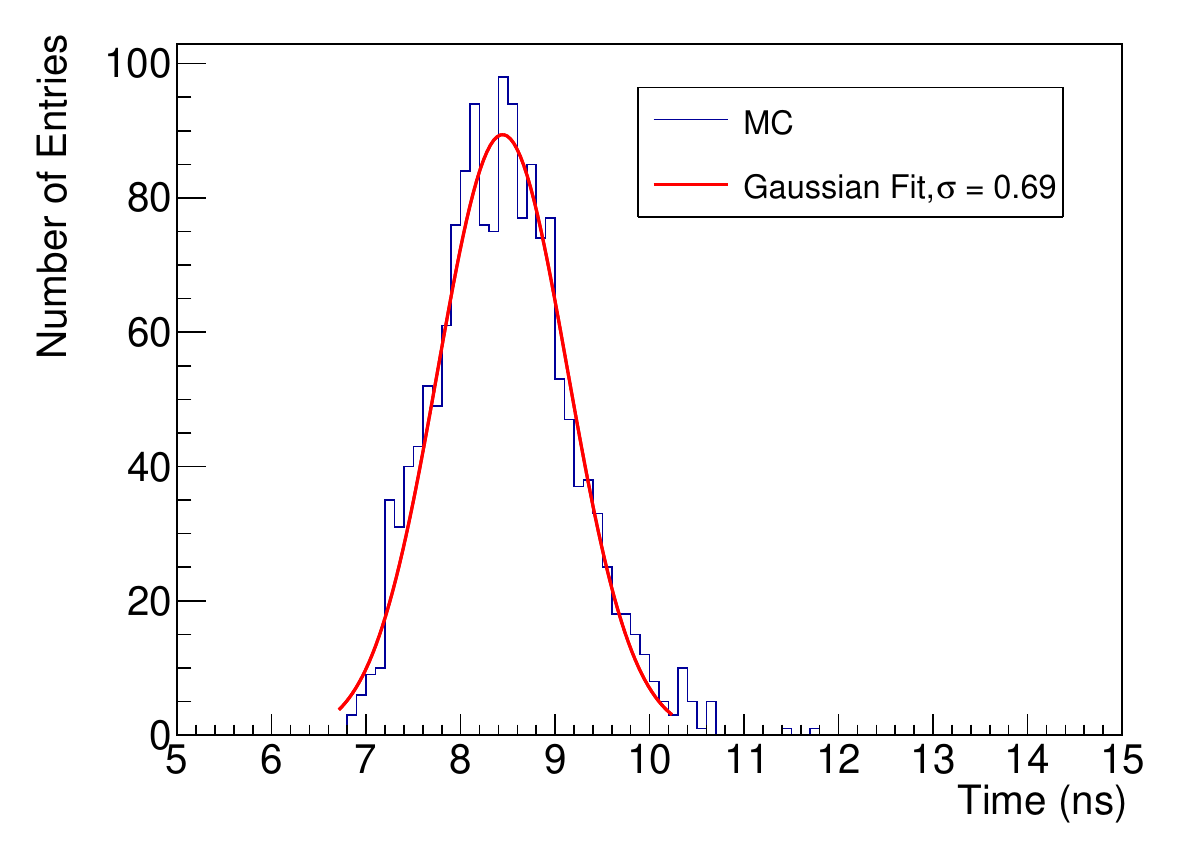}
    \caption{Simulated time distribution of a single SuperFGD channel (without electronics error). The standard deviations shows a time resolution of about 0.69 ns.}
    \label{fig:tr_MC}

\end{figure}

In this section, the measured time resolution derived from the CERN beam test data \cite{Blondel:2020sfgd}, further calibrated in \cite{TR_Alekseev:2022jki}, and
also described in Sec.~\ref{sec:2020_data},
was taken as the reference. 
We simulated 500 MeV muon source, selecting events with a light yield that exceeds 20 p.e., consistent with the analysis procedures described in \cite{TR_Alekseev:2022jki, Blondel:2020sfgd} (see Sec.~\ref{time_reso}). To be consistent with the measured data in \cite{Blondel:2020sfgd}, the simulated light yield was tuned to about 58 p.e./MIP/channel by changing directly the intrinsic light yield of the scintiillator.
To reproduce the setup of the beam test prototypes, in the simulation the event trigger was given by the time of the first photon detected by the MPPC. 
Figure.~\ref{fig:tr_MC} shows the scintillator time distribution, revealing a time resolution of 0.69 ns. The expected contribution from the SFGD electronic system is $2.5 \text{ ns}/\sqrt{12} = 0.7 \text{ ns}$ \cite{Blondel:2020sfgd}. Adding up all the contribution results in an overall time resolution of about 0.98 ns, which is consistent with the calibrated single channel time resolution of 0.97 ns in \cite{TR_Alekseev:2022jki}.

Additionally, we tested the correlation between the light yield and the time resolution, comparing with the measurement with a 266 nm laser data \cite{TR_Alekseev:2022jki} (see Sec.~\ref{time_reso}), close to the maximum absorption point of the first scintillator doping component (PTP).
The simulated time resolution as a function of the light yield was compared to the measured time resolution where the scintillator light yield could be controlled by changing the laser luminosity.
Similarly to Fig.~\ref{fig:tr_MC}, the standard deviation of the hit time distribution for a given light yield was taken as the corresponding time resolution.

\begin{figure}[h!]
    \centering
    \includegraphics[width=0.6\linewidth]{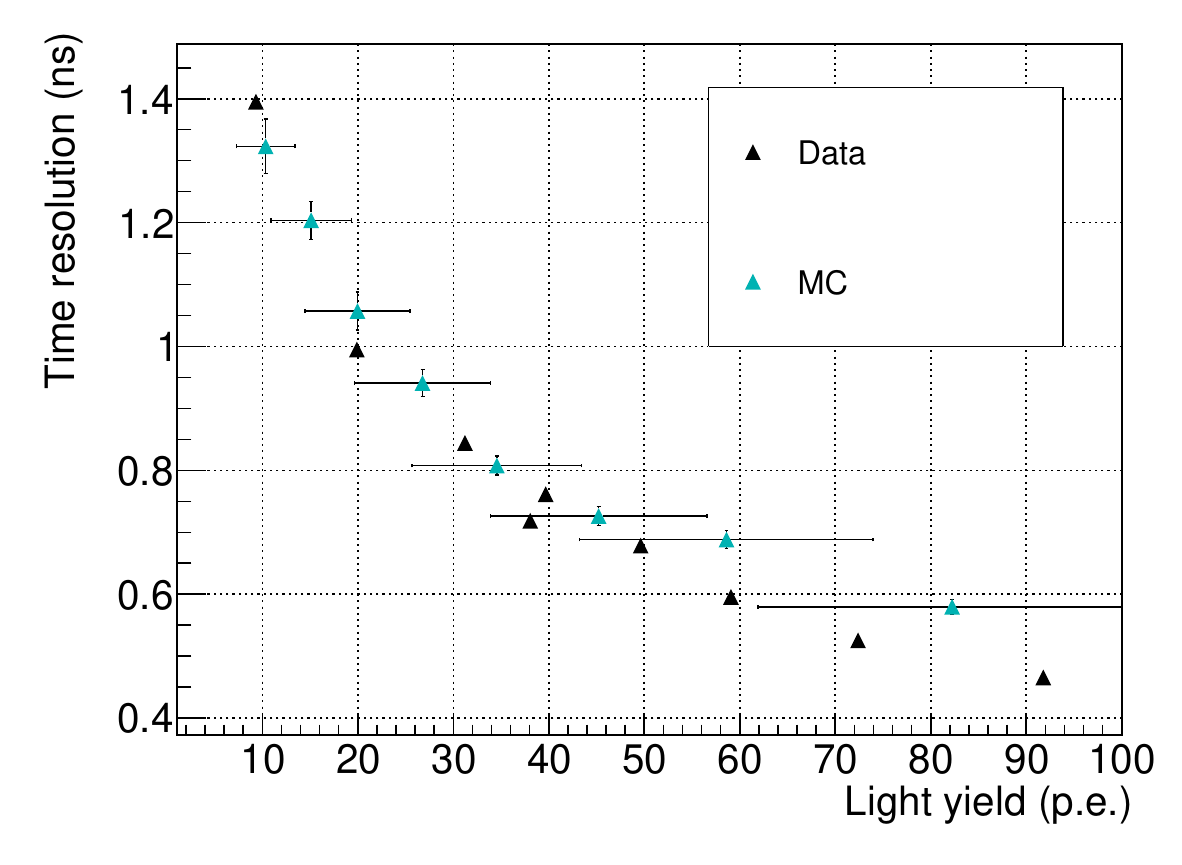}
    \caption{SuperFGD prototype single channel time resolution as a function of the average light yield. The black triangles represent the measured data using a 266 nm laser with various luminosity, and the blue triangles represent the muon simulation as a function of the plastic scintillator light yield. The data measurement is described in Sec.~\ref{time_reso}.}
    \label{fig:attcurve_vs_ly_MC}

\end{figure}

The results shown in Fig.~\ref{fig:attcurve_vs_ly_MC} demonstrate a good agreement between our simulation and the measured data. 
The energy deposition of a muon follows a Landau distribution, which resulted in the large variation in the detected light yield when the intrinsic scintillator light yield is high. This contributed to the deviation in time resolution between the data and the simulation in the high light yield region. Nevertheless, the nice data-MC matching within the typical light yield region below 60 p.e. validated the capability of the simulation to reproduce the measured time resolution of the SuperFGD.

\begin{figure}[h!]
    \centering

        \begin{subfigure}{0.3\textwidth}
            \includegraphics[width=\linewidth]{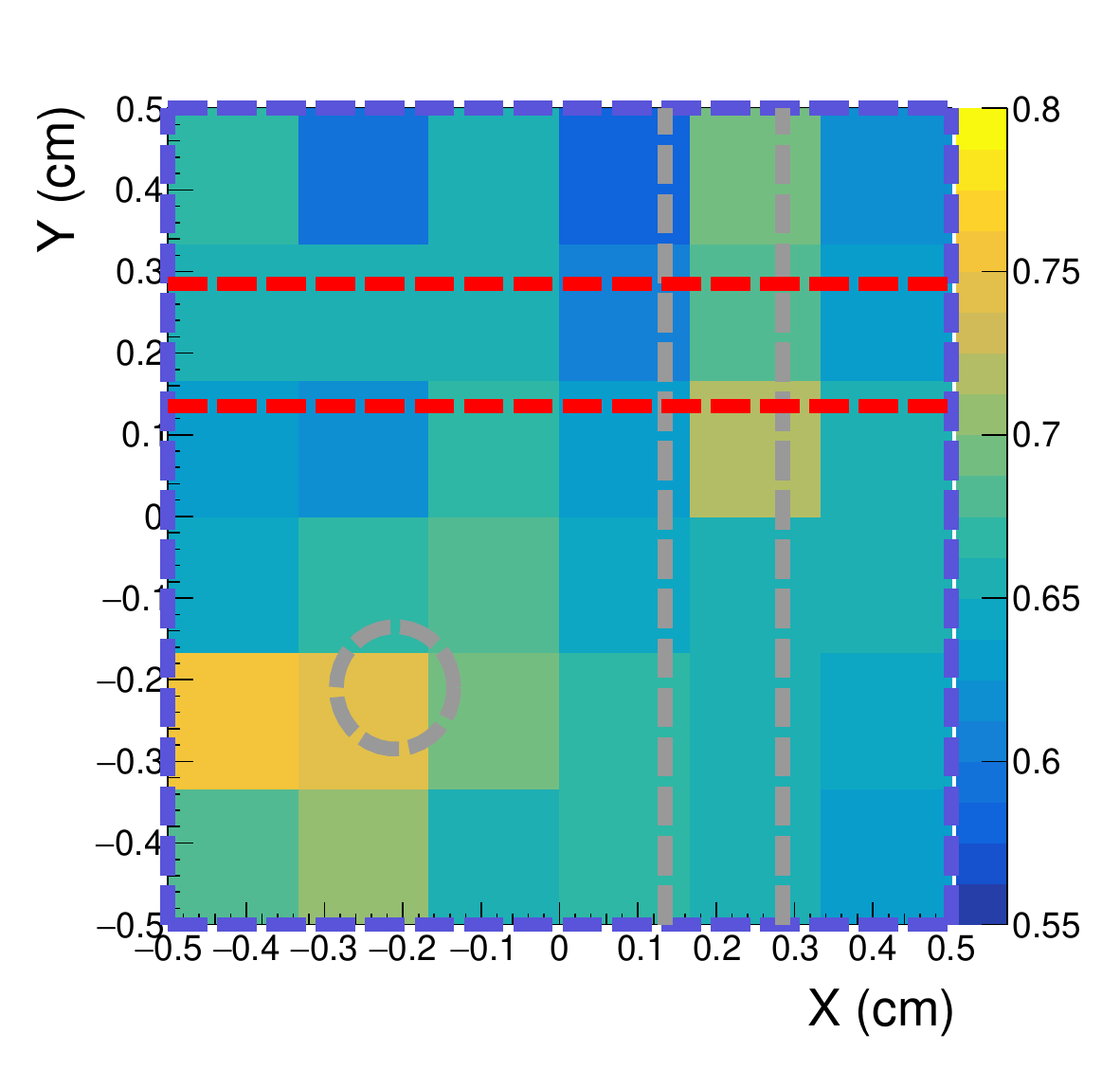}
            \caption{X Fiber readout}
        \end{subfigure}%
        \begin{subfigure}{0.3\textwidth}
            \includegraphics[width=\linewidth]{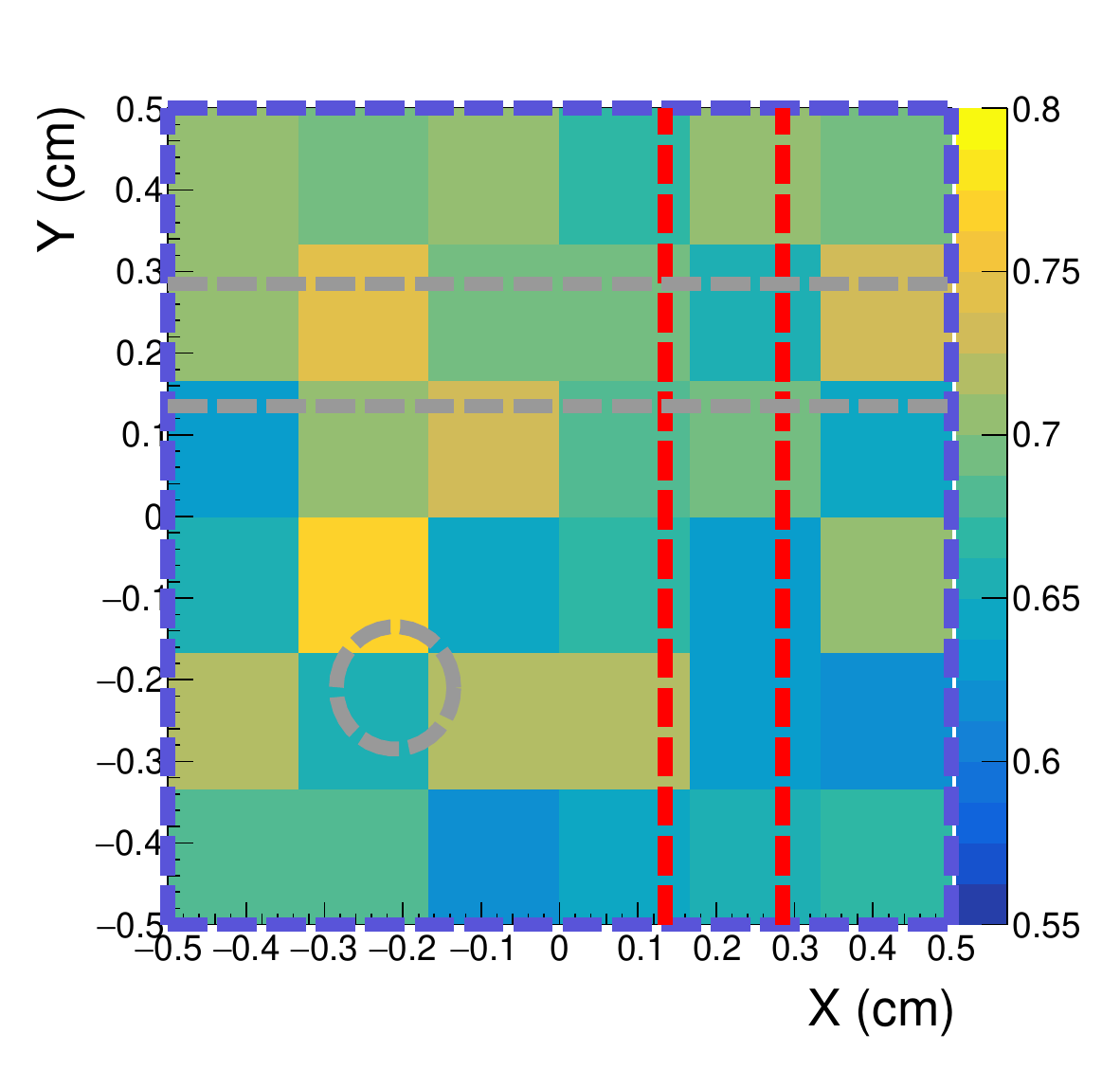}
            \caption{Y Fiber readout}
        \end{subfigure}
        \begin{subfigure}{0.3\textwidth}
            \includegraphics[width=\linewidth]{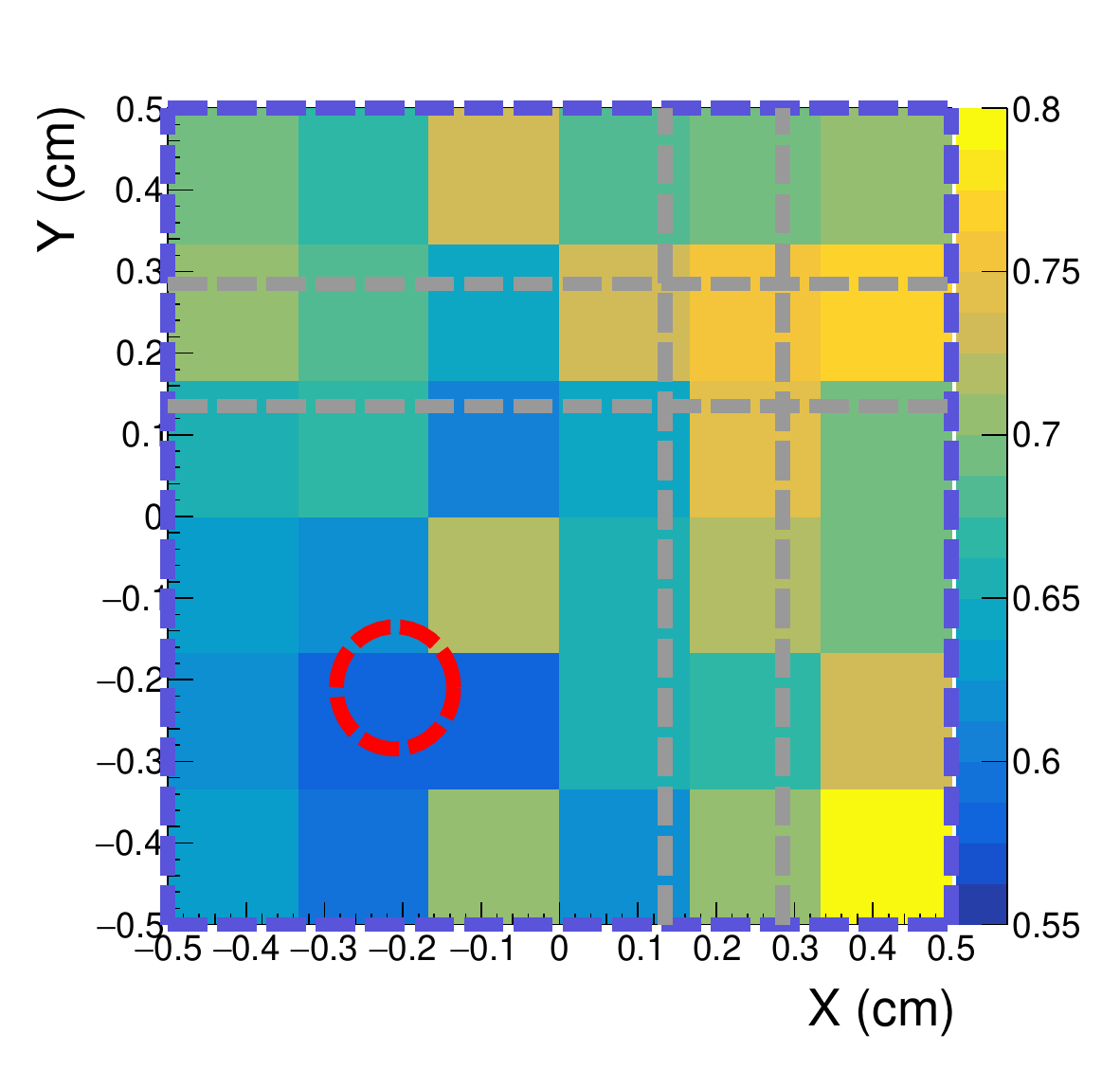}
            \caption{Z Fiber readout}
        \end{subfigure}%
        \caption{SuperFGD single-channel time resolution (ns) as a function of the particle position. 
        The X and Y fibers are perpendicular to the simulated particle track, while the Z fiber is along the track. The cube cross section was shown by blue dashed lines, while the position of the holes were shown in grey. The measuring fiber channel of each map is highlighted in red.}
        \label{fig:TRnonuni_MC}

\end{figure}

After validating the MC against data, optimized parameters were fixed as listed in Tab.~\ref{table:parameters}.
Then, we simulated the time resolution uniformity for a single SuperFGD cube as a function of particle position, 
which was not yet studied in the beam tests.
Positrons with 500 MeV kinetic energy were generated randomly along the z axis within an x-y plane segmented into $6\time6$ cells of size 0.16 $\times$ 0.16 $\text{mm}^2$ each. In each readout channel, cell-wise time resolution were obtained from the standard deviation of the corresponding time distributions of the detected hits. 
Figure.~\ref{fig:TRnonuni_MC} shows the simulated channel-wise time resolution uniformity map.

A similar pattern as the light yield non-uniformity maps was observed as expected, i.e., the time resolution is higher when the particle interacting position is closer to the WLS fibers, where a larger acceptance of the scintillating light leads to a higher light yield. 
The maximal spread in time resolution within a single cube is found to be about 300 ps, observed only when the hitting point of the traversing charged particle is close to the fired fiber, producing high lighy yield. The overall time resolution non-uniformity for each fiber channel, characterized by the standard deviation of the maps shown in Fig.~\ref{fig:TRnonuni_MC}, is 37 ps, 36 ps and 53 ps respectively, small compared to the total detector time resolution per channel, i.e., about 0.85 ns.

\section{Conclusions}
\label{sec:discussion}
In this study, a detailed optical simulation of the SuperFGD detector unit, an optically-isolated plastic scintillator cube crossed by three orthogonal WLS fibers, was developed. 
The optical model, implemented in Geant4, was tuned and validated on different datasets obtained by exposing different prototypes to charged particle beam tests as well as from measurements performed in different laboratories.
The optical simulation was proved to be able to reproduce well the response of the detector light yield as well as its uniformity as a function of the particle interacting position within the cube, optical crosstalk, the light attenuation in the WLS fiber and the time resolution. 
Eventually, the maximum time resolution non-uniformity, not available from the collected data, was simulated with the validated optical model and found to be negligible compared to the total SuperFGD time resolution.

Overall, given the validated reliability of the developed optical model of SuperFGD, it can be used to estimate more precisely the potential detector systematic uncertainties, and help to gain better understanding on the future SuperFGD neutrino data. As a delicate optical model, it possesses the potential to be extended and modified to be suitable of investigating the further upgrading of the segmented fine-granularity detectors.

\section*{Acknowledgements}

This work was supported in part in the framework of the State project ``Science'' by the Ministry of Science and Higher Education of the Russian Federation under the contract 075-15-2024-541. Part of this work was supported by the SNSF grant PCEFP2 203261, Switzerland.
We thank Research Center for Election Photon Science, Tohoku University for the allocation of beamtime (Proposal No. 2892 and 2911).

\section*{Author contributions}
\textbf{Botao Li}: Conceptualization, Methodology, Software, Validation, Data Curation, Writing - Original Draft, Visualization. 
\textbf{Davide Sgalaberna}: Supervision, Conceptualization, Writing-Original Draft, Writing - Review \& Editing. 
\textbf{Yury Kudenko}: Resources, Writing - Original Draft, Writing - Review \& Editing. 
\textbf{Tatsuya Kikawa}: Resources, Writing - Original Draft, Writing - Review \& Editing.
\textbf{All authors}: Review

%% If you have bib database file and want bibtex to generate the
%% bibitems, please use
%%
\bibliographystyle{JHEP} 
\bibliography{Optical_MC}

\end{document}